\documentclass[usenatbib]{mn2e}

\usepackage{mathrsfs,aas_macros}
\usepackage{color}
\usepackage{graphicx}
\usepackage{url}

\def\Lya{Ly$\alpha$~} 
\def\Lyb{Ly$\beta$~}

\def\LCDM{$\Lambda$CDM~}

\def\HI{\hbox{H$\,\rm \scriptstyle I\ $}}
 
\def\HeI{\hbox{He$\,\rm \scriptstyle I\ $}}
\def\HeII{\hbox{He$\,\rm \scriptstyle II\ $}}

\title[The Sherwood simulation suite]{The Sherwood simulation suite:
  overview and data comparisons with the Lyman-$\alpha$ forest at
  redshifts \boldmath{$2\leq z \leq 5$}}

\author[J.S. Bolton et al.] {James S. Bolton$^{1}$, Ewald
  Puchwein$^{2}$, Debora Sijacki$^{2}$, Martin G. Haehnelt$^{2}$,
  \newauthor Tae-Sun Kim$^3$, Avery Meiksin$^{4}$, John A. Regan$^{5}$
  \& Matteo Viel$^{3,6}$\\$^1$School of Physics and Astronomy,
  University of Nottingham, University Park, Nottingham, NG7 2RD,
  UK\\$^2$Kavli Institute for Cosmology and Institute of Astronomy,
  Madingley Road, Cambridge, CB3 0HA, UK\\$^3$INAF - Osservatorio
  Astronomico di Trieste, Via G.B. Tiepolo 11, I-34131 Trieste, Italy
  \\$^4$SUPA\thanks{Scottish Universities Physics Alliance}, Institute
  for Astronomy, University of Edinburgh, Blackford Hill, Edinburgh,
  EH9 3HJ, UK\\$^5$Institute for Computational Cosmology, Department
  of Physics, Durham University, South Road, Durham, DH1 3LE, UK
  \\$^6$INFN/National Institute for Nuclear Physics, Via Valerio 2,
  I-34127 Trieste, Italy }

\begin{document}

\date{\today}

\maketitle

\label{firstpage}

\begin{abstract}
We introduce a new set of large scale, high resolution hydrodynamical
simulations of the intergalactic medium: the Sherwood simulation
suite.  These are performed in volumes $10^{3}$--$160^{3}h^{-3}$
comoving $\rm Mpc^{3}$, span almost four orders of magnitude in mass
resolution with up to 17.2 billion particles, and employ a variety of
physics variations including warm dark matter and galactic
outflows. We undertake a detailed comparison of the simulations to
high resolution, high signal-to-noise observations of the \Lya forest
over the redshift range $2 \leq z \leq 5$.  The simulations are in
very good agreement with the observational data, lending further
support to the paradigm that the \Lya forest is a natural consequence
of the web-like distribution of matter arising in \LCDM cosmological
models.  Only a small number of minor discrepancies remain with
respect to the observational data.  Saturated \Lya absorption lines
with column densities $N_{\rm HI}>10^{14.5}\rm\,cm^{-2}$ at $2<z<2.5$
are underpredicted in the models.  An uncertain correction for
continuum placement bias is required to match the distribution and
power spectrum of the transmitted flux, particularly at $z>4$.
Finally, the temperature of intergalactic gas in the simulations may
be slightly too low at $z=2.7$ and a flatter temperature-density
relation is required at $z=2.4$, consistent with the expected effects
of non-equilibrium ionisation during \HeII reionisation.

\end{abstract}



\section{Introduction}

Hydrodynamical simulations of structure formation have convincingly
demonstrated that the \Lya forest is an excellent probe of the
underlying dark matter distribution, tracing the cosmic web of large
scale structure on scales $1$--$80h^{-1}$ comoving Mpc (cMpc) along
the line of sight
\citep{Cen1994,Zhang1995,Hernquist1996,MiraldaEscude1996,Theuns1998}.
Detailed comparison of intergalactic \Lya absorption line observations
to simulations at $2 \la z \la 5$ have placed constraints on the
matter power spectrum
\citep{Croft1999,McDonald2006,Viel2004,Palanque2013}, the ionisation
and thermal state of the intergalactic medium
\citep{Rauch1997,Dave1999,Schaye2000,MeiksinWhite2003,FaucherGiguere2008b,BeckerBolton2013},
the coldness of cold dark matter
\citep{Narayanan2000,Seljak2006,Viel2005,Viel2013}, and the baryon
acoustic oscillation scale \citep{Slosar2013,Busca2013}.

In the forthcoming decade, thirty metre class telescopes equipped with
high resolution ($R\geq 50\,000$) echelle spectrographs, coupled with
huge numbers of low-to-moderate resolution spectra ($R\sim
2000$--$5000$) obtained with proposed large scale quasar surveys with
the \emph{William Herschel Telescope Enhanced Area Velocity
  Explorer}\footnote{\url{http://www.ing.iac.es/weave/consortium.html}}
(WEAVE) and the \emph{Dark Energy Spectroscopic
  Instrument}\footnote{\url{http://desi.lbl.gov/}} (DESI), will open
up new vistas on the high redshift intergalactic medium (IGM) probed
by the \Lya forest.  These facilities and surveys will enable access
to fainter, more numerous background quasars, and will probe the IGM
transverse to the line of sight with densely packed background
sources.

Critical to all these programmes are high fidelity models of the IGM.
These are required for forward modelling the observational data and
facilitating (model-dependent) constraints on quantities of
cosmological and astrophysical interest.  A drawback of existing
hydrodynamical simulations of the IGM is their narrow dynamic range,
which translates to simulation volumes of $\sim
10^{3}$--$20^{3}h^{-3}\rm\, cMpc^{3}$ due to the requirement of
resolving structures with $M_{\rm gas} \sim 10^{6}h^{-1}M_{\odot}$
\citep{BoltonBecker2009} and spatial scales of $\sim 20h^{-1}\rm
\,ckpc$ \citep{Lukic2015}.  This requirement is problematic when
simulating correlations in the IGM on large scales, analysing the
properties of \Lya absorption systems around rare, massive haloes, and
correcting for the lack of large scale power in small simulation
volumes.  Convergence at the $<10$ per cent level in the \Lya forest
power spectrum requires gas particle masses of $\sim
10^{5}h^{-1}M_{\odot}$ and volumes $\geq 40^{3}h^{-3}\rm\,cMpc^{3}$ to
correctly capture the relevant large scale modes
\citep{Bryan1999,MeiksinWhite2001,McDonald2003,BoltonBecker2009,Tytler2009,Lidz2010,Borde2014,ArinyoiPrats2015,Lukic2015}.

In addition to achieving a sufficient dynamic range, it is important
to assess how faithfully the hydrodynamical simulations reproduce the
observational properties of the \Lya forest.  Although the overall
agreement between high resolution spectroscopic data at $2 \leq z \leq
5$ and \LCDM hydrodynamical simulations is astonishingly good, several
discrepancies have been highlighted over the last few decades.  These
include absorption line velocity widths in the simulations which are
too narrow \citep{Theuns1998,Bryan1999}, underdense gas that may be
significantly hotter than typically assumed in the models
\citep{Bolton2008,Viel2009}, and too few absorption lines in the
simulations with \HI column densities $10^{14}\rm\,cm^{-2}<N_{\rm
  HI}<10^{16}\rm\,cm^{-2}$ \citep{Tytler2009}.

In this work, we revisit these issues while introducing a new set of
state-of-the-art hydrodynamical simulations of the IGM -- the Sherwood
simulation suite -- performed with a modified version of the parallel
Tree-PM smoothed-particle hydrodynamics (SPH) code
\textsc{P-Gadget-3}, an updated and extended version of the publicly
available \textsc{Gadget-2} \citep{Springel2005}.  These are some of
the largest hydrodynamical simulations of the \Lya forest performed to
date, in volumes $10^{3}$--$160^{3}h^{-3}\rm\,cMpc^{3}$ with up to
$2\times 2048^{3}$ (17.2 billion) particles.  Other recent simulations
similar in scale include \Lya forest models performed with the
Eulerian \textsc{Nyx} hydrodynamical code \citep{Lukic2015}, as well
as the Illustris \citep{Vogelsberger2014} and EAGLE simulations
\citep{Schaye2015}.  The Sherwood simulations are similar in terms of
volume and resolution to the \Lya forest simulations presented by
\citet{Lukic2015}, but employ a different hydrodynamics algorithm and
explore a wider array of model parameters including warm dark matter
and galactic outflows.  Illustris and EAGLE -- performed with the
moving mesh code \textsc{Arepo} \citep{Springel2010} and a modified
version of \textsc{P-Gadget-3}, respectively -- have more
sophisticated sub-grid treatments for gas cooling, star formation and
feedback, but are performed at lower mass resolution.  The highest
resolution Sherwood runs furthermore stop at $z=2$, whereas Illustris
and EAGLE have been performed to $z=0$.

The Sherwood simulations are designed to fulfill a broad range of
roles, including studying the IGM during hydrogen reionisation at
$z\geq 6$, constraining the matter power spectrum with the \Lya forest
at $z\simeq 4$, mapping three dimensional structure in the IGM with
multiple quasars at $2<z<3$ and examining the properties of the low
redshift \Lya forest at $z<1$.  A subset of these models have already
been compared to metal line observations approaching reionisation at
$z \simeq 6$ \citep{Keating2016} and to the \Lya forest in an
ultra-high signal-to-noise spectrum ($\rm S/N\simeq 280$ per pixel) of
quasar HE0940$-$1050 at $z=3.09$ (Rorai et al. 2016, submitted).  In
this work -- the first in a series of papers -- we provide an overview
of the simulations and undertake a comparison to high resolution
observations of \Lya absorption at \HI column densities $N_{\rm
  HI}<10^{17.2}\rm\,cm^{-2}$ over the redshift range $2 \leq z \leq
5$, where observational data are readily accessible with optical
spectroscopy.  Our goal is not to advocate for a ``best fit'' set of
model parameters in this work. Rather, we instead report the
differences and agreements between our simulations and observational
data.  For a comprehensive set of reviews on this topic, we also refer
the interested reader to \citet{Rauch1998}, \citet{Meiksin2009} and
\citet{McQuinn2015}.

This paper is structured as follows.  We first describe the
simulations in Section~\ref{sec:method}.  In Section~\ref{sec:results}
we compare the simulations to a wide variety of observational data,
including the distribution and power spectrum of the transmitted flux,
the column density distribution function and the distribution of \Lya
absorption line velocity widths.  We summarise and present our
conclusions in Section~\ref{sec:conclude}.  The Appendix contains
numerical convergence tests.  We assume the cosmological parameters
$\Omega_{\rm m}=0.308$, $\Omega_{\Lambda}=0.692$ and $h=0.678$
throughout \citep{Planck2014} and refer to comoving and proper
distance units with the prefixes ``c'' and ``p'' respectively.


\section{Methodology}\label{sec:method}
\subsection{Hydrodynamical simulations}\label{sec:sims}

\begin{figure}
\begin{center}
  \includegraphics[width=0.47\textwidth]{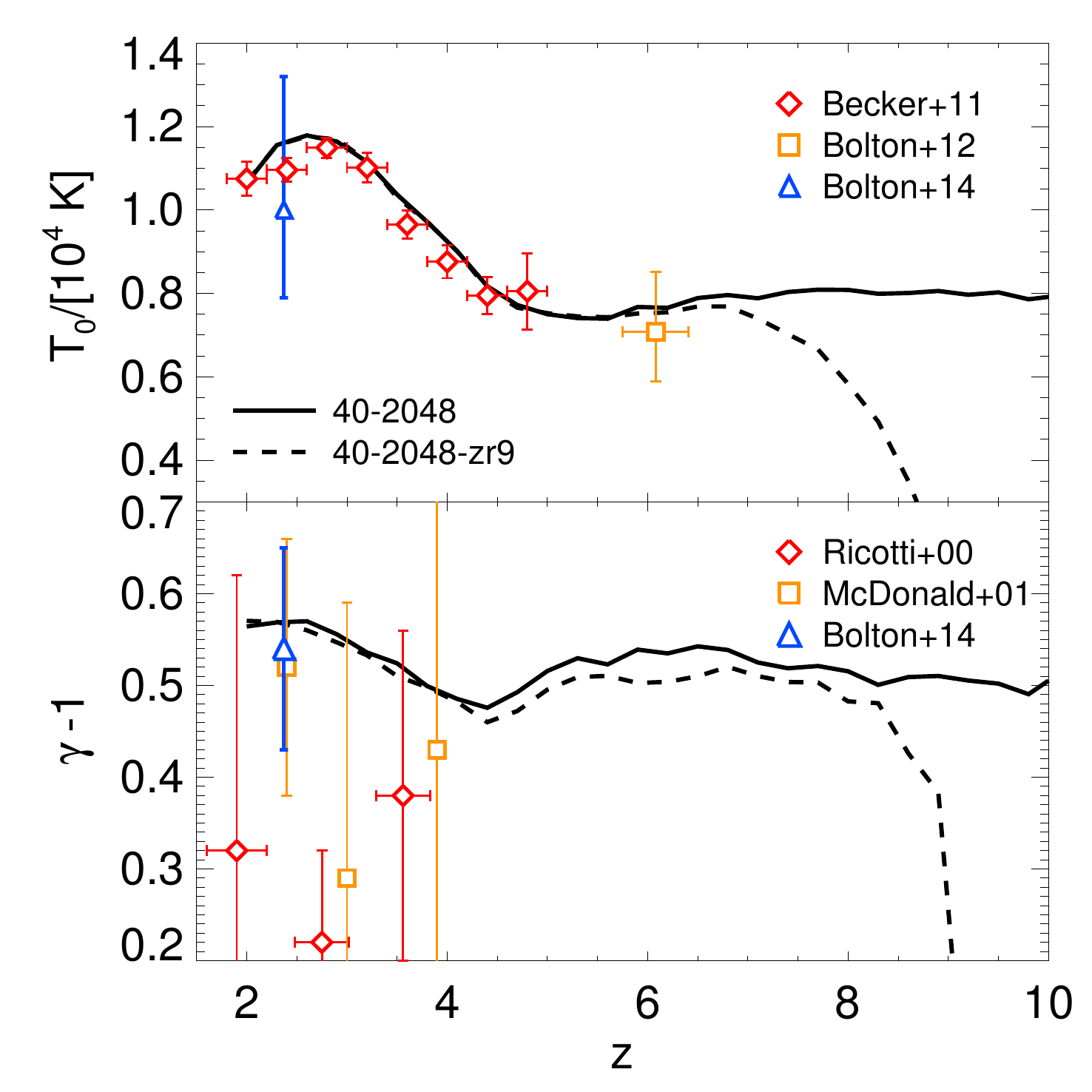}
  \vspace{-0.4cm}
  \caption{The two thermal histories used in the Sherwood
    simulations. The reference 40-2048 model corresponds to the
    \citet{HaardtMadau2012} ionising background (solid curve,
    reionisation at $z_{\rm r}=15$) with a small modification to the
    \HeII photo-heating rate (see text for details).  This is tuned to
    match observational constraints on the IGM temperature at mean
    density, $T_0$ (upper panel), obtained from the \Lya forest
    \citep{Becker2011,Bolton2012,Bolton2014}.  The 40-2048-zr9 model
    is similar at $z<6$, but with later reionisation at $z_{\rm
      r}=9$. The lower panel displays the power-law index, $\gamma-1$,
    of the temperature-density relation, $T=T_{0}\Delta^{\gamma-1}$,
    compared to \Lya forest measurements at $2<z<4$
    \citep{Ricotti2000,McDonald2001,Bolton2014}.  Note that
    \citet{Becker2011} measure the IGM temperature at the (redshift
    dependent) characteristic gas density probed by the \Lya forest,
    $T({\bar \Delta})$, rather than the temperature at mean density,
    $T_{0}$.  This characteristic density ranges from ${\bar
      \Delta}=1.2$--$5.7$ at $2\leq z \leq 5$ (see their table 3). The
    \citet{Becker2011} $T_{0}$ data points in the upper panel are thus
    obtained using the $\gamma-1$ evolution shown by the solid line in
    the lower panel, i.e. $T_{0}=T({\bar \Delta})/{\bar
      \Delta}^{\gamma-1}$.}
  \label{fig:Trho}
\end{center}
\end{figure}

\begin{table*}
  \centering
   \caption{Summary of the Sherwood simulation suite.  The models
     analysed in this work are listed in the upper part of the table.
     The naming convention in the first column, \emph{L-N-param},
     encodes the box size, $L$, in $\rm h^{-1}\,cMpc$, the cube root,
     $N$, of the gas particle number, and any parameters which have
     been varied from the reference run (40-2048).  These are: {\it
       wdm} --warm dark matter consisting of a $3.5\rm\, keV$ thermal
     relic \citep{Viel2013}; {\it zr9} -- rapid reionisation beginning
     at $z_{\rm r}=9$ (cf. $z_{\rm r}=15$ for the reference model);
     {\it ps13} -- models including a sub-resolution treatment for
     star formation and galactic outflows
     \citep{PuchweinSpringel2013}; {\it ps13+agn} -- as for {\it ps13}
     but with the addition of AGN feedback. Subsequent columns list
     the dark matter and gas particle masses in $h^{-1}M_{\odot}$, the
     gravitational softening length in comoving $h^{-1}\rm \,kpc$, the
     final redshift, $z_{\rm end}$, of the simulation, the choice of
     random seed for the initial conditions and comments on each
     model.}
  \begin{tabular}{|c|c|c|c|c|c|c|c|}
    \hline
    \hline
    Name & $M_{\rm dm}$ & $M_{\rm gas}$ & $l_{\rm soft}$ & $z_{\rm end}$ & Seed & Comments\\
         & $\rm [h^{-1}\,M_{\odot}]$ & $\rm [h^{-1}\,M_{\odot}]$ & $\rm\,[h^{-1}\,ckpc]$ & & & \\
    \hline
    {\bf 40-2048} & \boldmath{$5.37\times 10^{5}$} & \boldmath{$9.97\times 10^{4}$} & {\bf 0.78} & {\bf 2}  & A & {\bf Reference model}\\
    40-2048-wdm  & $5.37\times 10^{5}$ & $9.97\times 10^{4}$ & 0.78 & 2 & A & Warm dark matter, $3.5\rm\,keV$ thermal relic\\
    40-2048-zr9   & $5.37\times 10^{5}$ & $9.97\times 10^{4}$ & 0.78 & 2 & A & Ionising background at $z\leq9$ only\\
 
    80-2048       & $4.30\times 10^{6}$ & $7.97\times 10^{5}$ & 1.56 & 2 & B &\\
    40-1024       & $4.30\times 10^{6}$ & $7.97\times 10^{5}$ & 1.56 & 2 & A &\\
    40-1024-ps13  & $4.30\times 10^{6}$  & $7.97\times 10^{5}$ & 1.56 & 2 & A &\citet{PuchweinSpringel2013} winds \\
    20-512        & $4.30\times 10^{6}$ & $7.97\times 10^{5}$ & 1.56 & 2 & C & \\  
    40-512        & $3.44\times 10^{7}$ & $6.38\times 10^{6}$ & 3.13 & 0 & A  &\\
    \hline
    40-2048-ps13  & $5.37\times 10^{5}$ & $9.97\times 10^{4}$ & 0.78 & 5.2  & A &\citet{PuchweinSpringel2013} winds \\
    20-1024       & $5.37\times 10^{5}$ & $9.97\times 10^{4}$ & 0.78 & 2 & C &\\
    10-512        & $5.37\times 10^{5}$ & $9.97\times 10^{4}$ & 0.78 & 2 & A  &\\
  160-2048      & $3.44\times 10^{7}$ & $6.38\times 10^{6}$ & 3.13 & 2 & C &\\
    80-1024       & $3.44\times 10^{7}$ & $6.38\times 10^{6}$ & 3.13 & 0 & B &\\
    20-256        & $3.44\times 10^{7}$ & $6.38\times 10^{6}$ & 3.13 & 0 & C &\\
    160-1024      & $2.75\times 10^{8}$ & $5.10\times 10^{7}$ & 6.25 & 2 & C &\\
    80-512        & $2.75\times 10^{8}$ & $5.10\times 10^{7}$ & 6.25 & 0 & B &\\
    80-512-ps13   & $2.75\times 10^{8}$ & $5.10\times 10^{7}$ & 6.25 & 0 & B &\citet{PuchweinSpringel2013} winds \\
   80-512-ps13+agn & $2.75\times 10^{8}$ & $5.10\times 10^{7}$ & 6.25 & 0 & B &\citet{PuchweinSpringel2013} winds and AGN \\
   160-512       & $2.20\times 10^{9}$ & $4.08\times 10^{8}$ & 12.50 & 2 & C &\\
   80-256        & $2.20\times 10^{9}$ & $4.08\times 10^{8}$ & 12.50 & 0 & B &\\

    \hline
    \hline
  \end{tabular}

  \label{tab:sims}
\end{table*}

The Sherwood simulations are summarised in Table~\ref{tab:sims}.
These consist of 20 models that span almost four orders of magnitude
in mass resolution and employ a variety of physics parameters.  The
simulations required a total of 15 million core hours to run on the
\emph{Curie}
supercomputer\footnote{\url{http://www-hpc.cea.fr/en/complexe/tgcc-curie.htm}}
in France at the \emph{Tr{\'e} Grand Centre de Calcul}, using up to
6912 cores per run. The eight models analysed in this work are listed
in the upper part of the table -- the other models will be described
in detail elsewhere \citep[see e.g.][]{Keating2016}.

All the simulations were performed with a modified version of the
parallel Tree-PM SPH code \textsc{P-Gadget-3}, which is an updated and
extended version of the publicly available \textsc{Gadget-2}
\citep{Springel2005}.  The models use the best fit $\Lambda$CDM {\it
  Planck+WP+highL+BAO} cosmological parameters \citep{Planck2014},
where $\Omega_{\rm m}=0.308$, $\Omega_{\Lambda}=0.692$, $h=0.678$,
$\Omega_{\rm b}=0.0482$, $\sigma_{8}=0.829$ and $n=0.961$.  A
primordial helium fraction by mass of $Y_{\rm p}=0.24$ is assumed
throughout.  The simulations do not include metal line cooling.  This
is expected to have very little effect on the \Lya forest transmission
as the metallicity of the low density IGM is very low
\citep{Viel2013feedback}.  Initial conditions were generated at $z=99$
on a regular grid using the \textsc{N-GenIC} code
\citep{Springel2005Nat} using transfer functions generated by
\textsc{CAMB} \citep{Lewis2000}.  The single exception is the
40-2048-wdm model, where the linear matter power spectrum has been
suppressed at small scales to correspond to a warm dark matter thermal
relic with mass $3.5\rm\,keV$, consistent with lower limits inferred
from the \Lya forest at $z>4$ \citep{Viel2013}. The gravitational
softening length is set to $1/25^{\rm th}$ of the mean interparticle
spacing and the SPH kernel uses $64$ neighbour particles in all
simulations.  Three different random seeds were used to generate the
initial conditions.  The same seed was used for simulations with the
same box size (see Table~\ref{tab:sims}) so that the same large scale
structures are present.

The photo-ionisation and photo-heating of the hydrogen and helium gas
is calculated using the spatially uniform \citet{HaardtMadau2012}
ionising background model.  The gas is assumed to be optically thin
and in ionisation equilibrium.  Except in rare, highly overdense
regions, this is expected to be a very good approximation for the
post-reionisation IGM at $z\leq 5$, when hydrogen is highly ionised
and the mean free path for ionising photons is significantly larger
than the typical separation of ionising sources
\citep[e.g.][]{Worseck2014}.  However, inhomogeneous \HeII
reionisation may affect the gas temperatures at $z<4$, which is
important for line widths but of secondary importance to the
transmitted \Lya flux (we discuss this further in
Section~\ref{sec:bdist}).  The ionisation fractions are obtained
following the method outlined by \citet{Katz1996}, but with the case-A
recombination rates from \citet{VernerFerland1996}, the dielectric
\HeI recombination rate from \citet{Aldrovandi1973}, collisional
ionisation rates from \citet{Voronov1997}, collisional excitation
cooling rates from \citet{Cen1992}, the bremsstrahlung cooling rate
from \citet{Theuns1998} and the inverse Compton cooling rate from
\citet{Weymann1966}, for free electrons scattering off cosmic
microwave background photons with temperature $T_{\rm
  CMB}=2.73\rm\,K\, (1+z)$.  A small modification to the \HeII
photo-heating rate, $\epsilon_{\rm HeII}=1.7\epsilon_{\rm HeII}^{\rm
  HM12}$ for $2.2<z<3.4$, is applied to match observational
measurements of the IGM temperature at mean density, $T_0$, inferred
from the curvature of \Lya forest absorption lines \citep{Becker2011}.
A boost to the IGM temperature from non-equilibrium and radiative
transfer effects during \HeII reionisation is expected at these
redshifts \citep{AbelHaehnelt1999,McQuinn2009,Puchwein2015}.

The \citet{HaardtMadau2012} model furthermore results in the IGM being
quickly reionised at $z_{\rm r}=15$.  The most important effect this
choice has on the \Lya forest at $z<5$ is to alter the pressure
smoothing scale of gas in the simulations
\citep[e.g.][]{GnedinHui1998,Pawlik2009,Kulkarni2015}; gas which has
been ionised and heated later will have had less time to dynamically
respond to the change in pressure.  We thus also consider an
alternative model, 40-2048-zr9, where the \citet{HaardtMadau2012}
background has been modified to reionise the IGM at $z_{\rm r}=9$ but
retains a similar evolution in the gas temperature at $z<6$.  This
later reionisation model is in better agreement with the latest
measurement of the Thomson scattering optical depth for cosmic
microwave background photons, which is consistent with instantaneous
reionisation at $z_{\rm r}=8.8^{+1.7}_{-1.4}$ \citep{Planck2015}. A
comparison of the reference (40-2048) and 40-2048-zr9 thermal
histories to observational measurements of the power-law IGM
temperature-density relation, $T=T_{0}\Delta^{\gamma-1}$, where
$\Delta=\rho/\langle \rho \rangle$ is the gas density relative to the
background density \citep{HuiGnedin1997,McQuinn2016}, is displayed in
Fig.~\ref{fig:Trho}. We note that a fully self-consistent treatment of
pressure smoothing would require radiation hydrodynamical simulations
that model patchy reionisation \citep[e.g.][]{Pawlik2015}, but these
are too computationally expensive at present for modelling the $2 \leq
z \leq 5$ \Lya forest at high resolution.  However, we do not expect
these differences to be large, particularly at $z<5$; thermal pressure
dominates over any hydrodynamical response of the gas to passing
cosmological ionisation fronts, which can have velocities approaching
a significant fraction of the speed of light \citep[see
  also][]{Finlator2012}.  The dynamical effects of \HeII reionisation
at $ z\simeq 3$ are also at the few per cent level in dense regions
when compared to simulations performed in the optically thin limit
\citep{MeiksinTittley2012}.

Star formation is not followed in most of the simulations.  Instead,
gas particles with temperature $T<10^{5}\rm\,K$ and an overdensity
$\Delta>1000$ are converted to collisionless particles
\citep{Viel2004}, resulting in a significant increase in computation
speed at the expense of removing cold, dense gas from the model.  This
choice has a minimal effect on the low column density absorption
systems probed by the \Lya forest.  We do, however, perform some
simulations with the star formation and energy-driven outflow model of
\citet{PuchweinSpringel2013} to investigate this further.  The star
formation prescription in this model follows
\citet{SpringelHernquist2003}, although it assumes a Chabrier rather
than a Salpeter initial mass function. This increases the available
supernovae feedback energy by a factor of $\sim 2$.  Furthermore,
rather than assuming a constant value of the wind velocity $v_w$, it
scales with the escape velocity of the galaxy.  The mass outflow rate
in units of the star formation rate is then computed from the wind
velocity and the available energy, i.e. it scales as $~v_w^{-2}$. This
increased mass loading in low mass galaxies results in a much better
agreement with the observed galaxy stellar mass function below the
knee at low redshift, as well as in a suppression of excessive
high-redshift star formation. Good agreement with the observed
$z\sim6$ galaxy stellar mass function is obtained \citep{Keating2016}.
The wind and star formation model parameters are the same as in the
S15 run of \citet{PuchweinSpringel2013}.  The simulations discussed in
this work do not include feedback from active galactic nuclei (AGN),
although we have performed one low resolution run (80-512-ps13+agn)
with the AGN feedback model described in \citet{PuchweinSpringel2013}
for examining the \Lya forest at $z<2$.  AGN feedback is expected to
impact on \Lya forest transmission statistics at the $5$--$10$ per
cent level at $z=2.25$, but has a substantially smaller effect on the
\Lya forest at higher redshift \citep{Viel2013feedback}.

Finally, we note that \Lya forest statistics are predominantly
sensitive to low density gas ($\Delta \la 10$), and differ at only the
$\sim 5$ per cent level between ``standard'' SPH used in this work and
moving-mesh \citep{Bird2013} or grid based codes \citep{Regan2007}.
Uncertainties associated with high resolution observational
measurements of the \Lya forest are typically similar to or larger
than this, and at present this is not expected to significantly alter
our conclusions.

\subsection{Mock \Lya forest spectra}\label{sec:spectra}

\begin{figure*}
  \begin{minipage}{0.73\textwidth}
    \includegraphics[width=\textwidth]{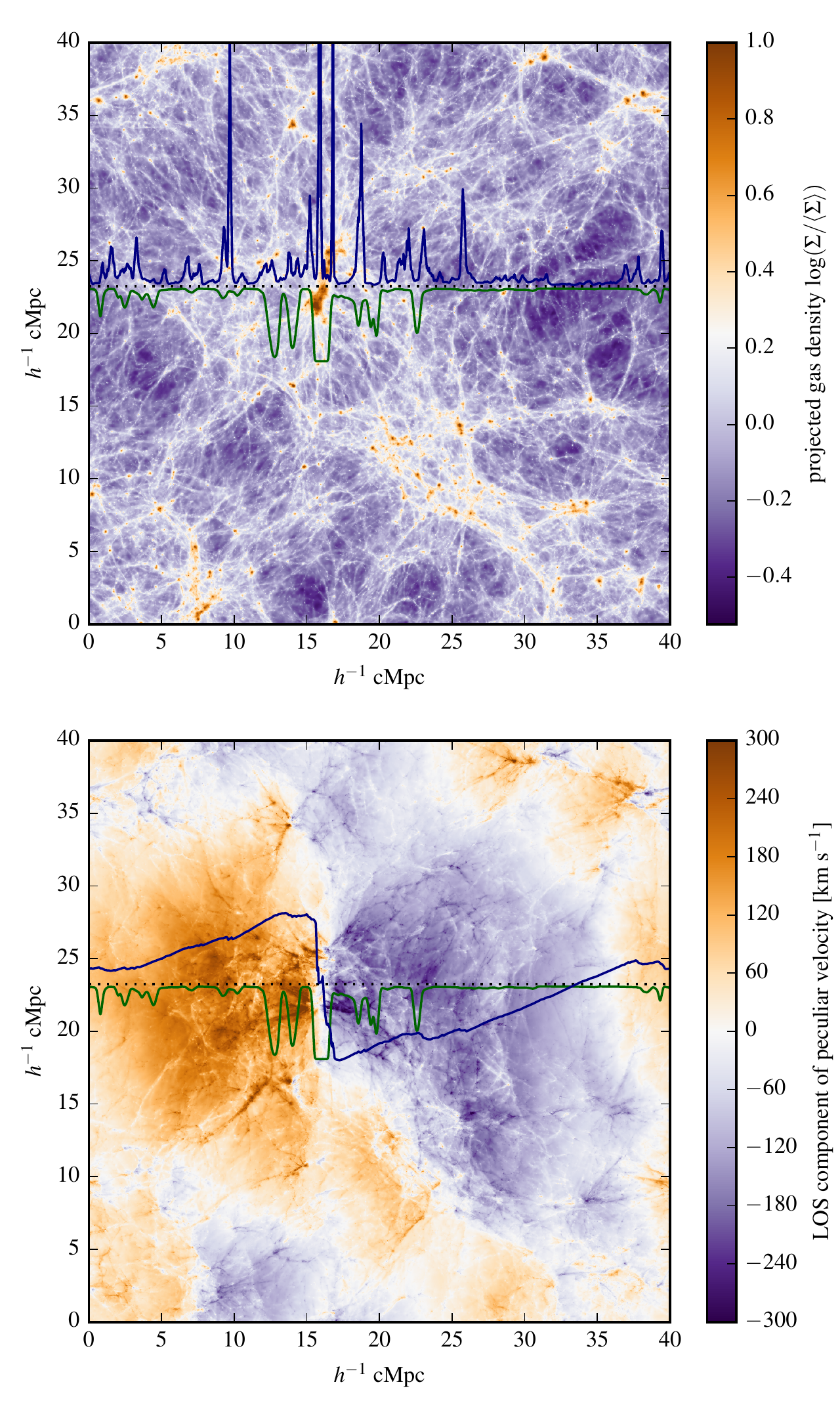}
  \end{minipage}
 \vspace{-0.6cm}
  \caption{Projections of the proper gas density (upper panel) and the
    component of the peculiar velocity along the horizontal axis
    (lower panel) in the y-z plane of the 40-2048-zr9 simulation at
    $z=2$.  The projection is performed along the entire box length of
    $40h^{-1}\,\rm cMpc$.  Neutral hydrogen that traces the web-like
    distribution of gas in the upper panel is responsible for the
    absorption lines observed in the \Lya forest.  This is illustrated
    by the horizontal dotted line, which marks the location of the
    line of sight corresponding to the \Lya absorption spectrum shown
    in green.  The superimposed blue curves display the corresponding
    line of sight gas density (upper panel) and peculiar velocity
    (lower panel).  Note the peculiar velocities shown by the blue
    curve are scaled such that the distance from the dotted line
    corresponds to the shift between real and redshift space along the
    line of sight.}
  \label{fig:projection}  
\end{figure*}

Mock \Lya absorption spectra were extracted on-the-fly at redshift
intervals of $\Delta z=0.1$ from all simulations.  At each redshift,
5000 lines of sight with 2048 pixels were drawn parallel to each of
the box axes ($x$, $y$ and $z$) on regularly spaced grids of $50^{2}$,
$40^{2}$ and $30^{2}$, respectively.  The gas density, \HI fraction,
\HI weighted temperature and \HI weighted peculiar velocities were
extracted following \citet{Theuns1998}.  The \HI \Lya optical depth
along each line of sight, $\tau_{\rm HI}^{\alpha}$, was computed using
the Voigt profile approximation provided by \citet{TepperGarcia2006}.
The transmitted flux in each pixel is then $F=e^{-\tau_{\rm
    HI}^{\alpha}}$.
 
An example mock \Lya forest spectrum and projections of the gas
density and the component of the peculiar velocity along the
horizontal axis in the 40-2048-zr9 model at $z=2$ are displayed in
Fig.~\ref{fig:projection}.  The \Lya absorption lines arise from
mildly overdense regions at this redshift, although the absorption
features are typically offset from the physical location of the gas.
This is because peculiar velocity gradients associated with structure
formation impact on the location of the \Lya absorption features in
velocity space.  For example, the high density peaks at $10h^{-1}\rm\,
cMpc$ and $25h^{-1}\rm\, cMpc$ along the horizontal axis correspond to
the absorption features at approximately $12h^{-1}\rm\, cMpc$ and
$23h^{-1}\rm\, cMpc$.  These density peaks are associated with the
filaments in the simulation volume.  The most massive halo is located
near the centre of the gas density projection in the upper panel of
Fig.~\ref{fig:projection}, with a dark matter mass of $2.4\times
10^{13}h^{-1}M_{\odot}$.  The component of the peculiar velocity along
the line of sight clearly shows gravitational infall around this high
density region; positive (orange) peculiar velocities indicate gas
that is moving from left to right, whereas negative (blue) velocities
show gas that is moving from right to left.

Further comparison of the simulations to observations requires
matching the properties of the observational data as closely as
possible \citep{Rauch1997,Meiksin2001}.  Throughout this paper we
follow the standard practice of rescaling the optical depth in each
pixel of the mock spectra by a constant factor to match the observed
evolution of the \Lya forest effective optical depth, $\tau_{\rm
  eff}(z)=-\ln \langle F \rangle$
\citep{Theuns1998,Bolton2005,Lukic2015}.  Here $\langle F
\rangle=\langle f_{\rm obs}/C_{\rm est} \rangle$ is the mean \Lya
forest transmission, $f_{\rm obs}$ is the observed flux in a
resolution element of the background quasar spectrum and $C_{\rm est}$
is the corresponding estimate for the continuum level.

There are a wide range of effective optical depth measurements quoted
in the literature
\citep[e.g.][]{Kim2002,Bernardi2003,Schaye2003,FaucherGiguere2008,Paris2011,Becker2013}.
In this work, when comparing simulations directly to published
observational data sets we therefore always rescale to match the
$\tau_{\rm eff}$ quoted in the publication where the observations were
first presented.  If we instead are comparing one simulation to
another without reference to any observational measurements, we then
adopt the following form for the redshift evolution of $\tau_{\rm
  eff}$ \citep{Viel2013}:

\begin{eqnarray}
\tau_{\rm eff}(z)  = \left\{ \begin{array}{cl} 
    1.274\left( \frac{1+z}{5.4} \right)^{2.90} - 0.132, & 2 \le z \le 4.4,\\ 
     1.142\left( \frac{1+z}{5.4} \right)^{4.91}, & 4.4 < z \le 5.
\end{array} \right.
\label{eq:taueff}
\end{eqnarray}

\begin{figure}
\begin{center}
  \includegraphics[width=0.47\textwidth]{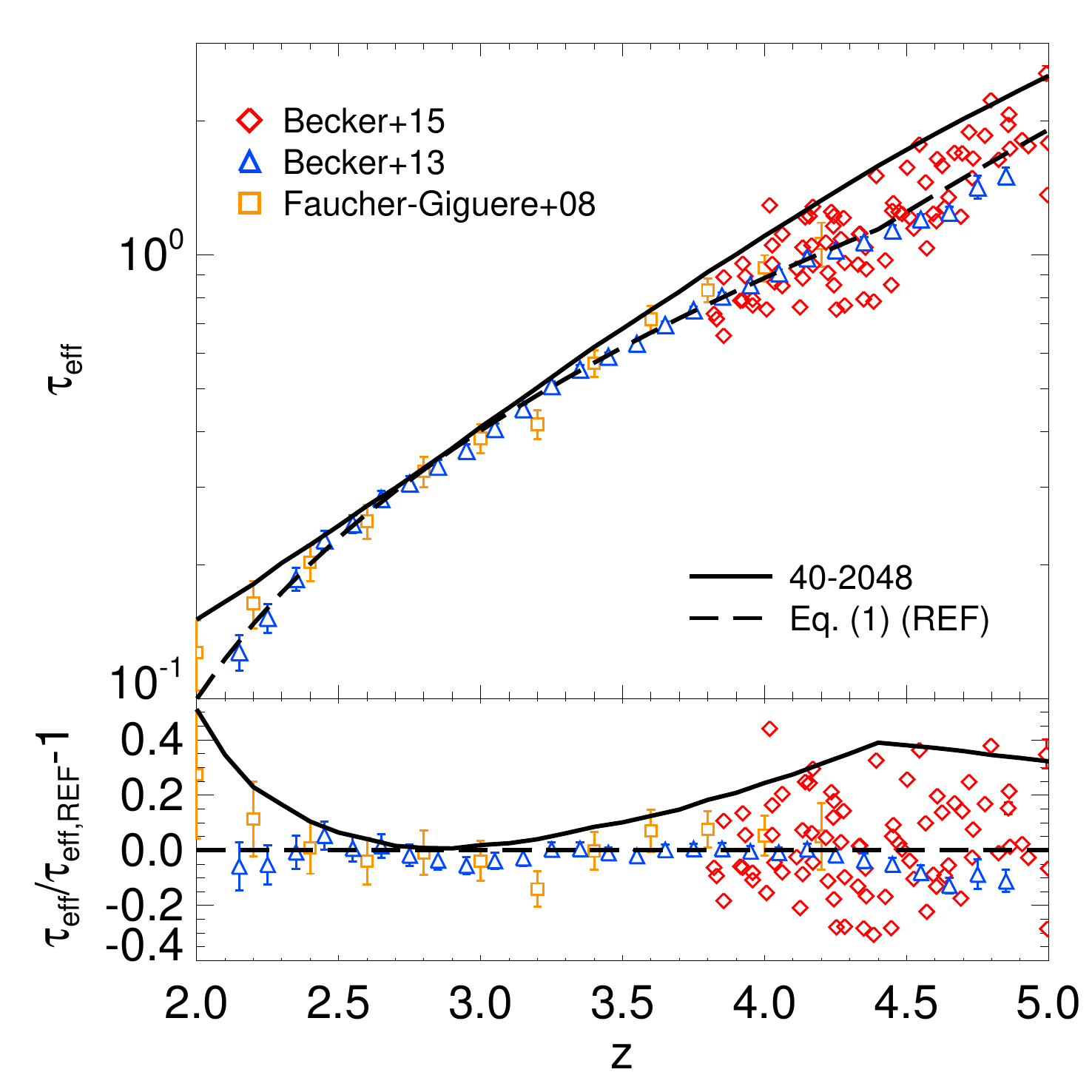}
  \vspace{-0.3cm}
  \caption{{\it Upper panel:} Observational measurements of the \Lya
    forest effective optical depth with redshift from
    \citet{FaucherGiguere2008} and \citet{Becker2013,Becker2015}.  The
    solid curve shows the redshift evolution in the 40-2048
    simulation, obtained from mock spectra with a total path length of
    $10^{5}h^{-1}\rm\,cMpc$. The dashed curve, which more closely
    represents the data, is given by Eq.~(\ref{eq:taueff}).  {\it
      Lower panel:} The simulated and observational data relative to
    Eq.~(\ref{eq:taueff}).}
  \label{fig:taueff}
\end{center}
\end{figure}

\noindent
The fit at $z \leq 4.4$ corresponds to eq. (5) in \citet{Becker2013},
with a steeper redshift evolution at $4.4< z < 5$ similar to
\citet{Fan2006}.  A comparison of Eq.~(\ref{eq:taueff}) to the
unscaled $\tau_{\rm eff}(z)$ from the reference model and
observational data is displayed in Fig.~\ref{fig:taueff}.  As already
pointed out by \citet{Puchwein2015}, the \citet{HaardtMadau2012}
ionising background overestimates $\tau_{\rm eff}$ at $2<z<2.5$ and
$z>4$.  In order for the 40-2048 model to match Eq.~(\ref{eq:taueff}),
we find the \HI photo-ionisation rate, $\Gamma_{\rm HI}$, predicted by
the \citet{HaardtMadau2012} model must be increased by
$[87,\,3,\,47,\,65]$ per cent at $z=[2,\,3,\,4,\,5]$, corresponding to
$\Gamma_{\rm HI}=[1.76,\,0.86,\,0.83,\,0.71]\times
10^{-12}\rm\,s^{-1}$.  For a more detailed study of the \HI
photo-ionisation rate at $2<z<5$ that includes an analysis of
systematic uncertainties, see \citet{BeckerBolton2013}.

Once rescaled to match the appropriate $\tau_{\rm eff}$, all mock
spectra are convolved with a Gaussian instrument profile with a Full
Width Half Maximum (FWHM) of $7\rm\,km\,s^{-1}$, typical of high
resolution ($R\sim 40\,000$) \Lya forest observations obtained with
echelle spectrographs.  The spectra are rebinned onto $3
\rm\,km\,s^{-1}$ pixels and uniform, Gaussian distributed noise is
added.  We adopt a signal-to-noise ratio of $\rm S/N=50$ per pixel
unless otherwise stated, typical of the high resolution data sets we
compare to.

The final adjustment we make to the mock spectra is a correction to
the effective optical depth arising from systematic bias in the
continuum level, $C_{\rm est}$, estimated in the observational data.
This correction also changes the shape of the transmitted flux
distribution and \Lya absorption lines.  Identifying the continuum
level in the \Lya forest becomes challenging toward higher redshift as
the effective optical depth increases.  The continuum may be placed
too low on the observational data if the \Lya forest is optically
thick, and there can also be large scale variations in the continuum
placement along individual spectra.  \citet{FaucherGiguere2008}
quantify this bias by manually fitting the continuum, $C_{\rm est}$,
on mock \Lya forest spectra where the true continuum, $C_{\rm true}$,
has been deliberately hidden. This yields an estimate for a continuum
correction $C_{\rm corr} = C_{\rm est}/C_{\rm true}$.  In this work,
unless otherwise stated we follow \citet{FaucherGiguere2008} and adopt
their estimate for the mean continuum error at $2\leq z \leq 4$

\begin{equation} 
 C_{\rm corr}(z) = 1-1.58\times10^{-5}(1+z)^{5.63}.
\label{eq:Ccorr}
\end{equation}


\noindent
As \citet{FaucherGiguere2008} do not determine $C_{\rm corr}$ at
$z>4.5$, when comparing to observational data at $z=5$ in this work we
instead follow \citet{Viel2013} and estimate a maximum correction of
$C_{\rm corr}=0.8$.  We forward model this continuum bias by applying
$C_{\rm corr}$ to mock spectra where the ``true'' continuum is already
known.  We use an iterative procedure where the optical depths in each
pixel are first rescaled to match the observed $\tau_{\rm eff}$, the
resolution and noise properties of the mock spectra are adjusted and
finally the continuum bias correction is applied as $ e^{-\tau_{\rm
    HI}^{\alpha}}/C_{\rm corr}$, again for each pixel \citep[see
  also][]{Rauch1997,Meiksin2001}.  This procedure is repeated until
convergence on the required $\tau_{\rm eff}$ is achieved.  Results
from mock spectra where this procedure has been applied have labels
appended with ``cc'' throughout.

\begin{figure}
\begin{center}
  \includegraphics[width=0.47\textwidth]{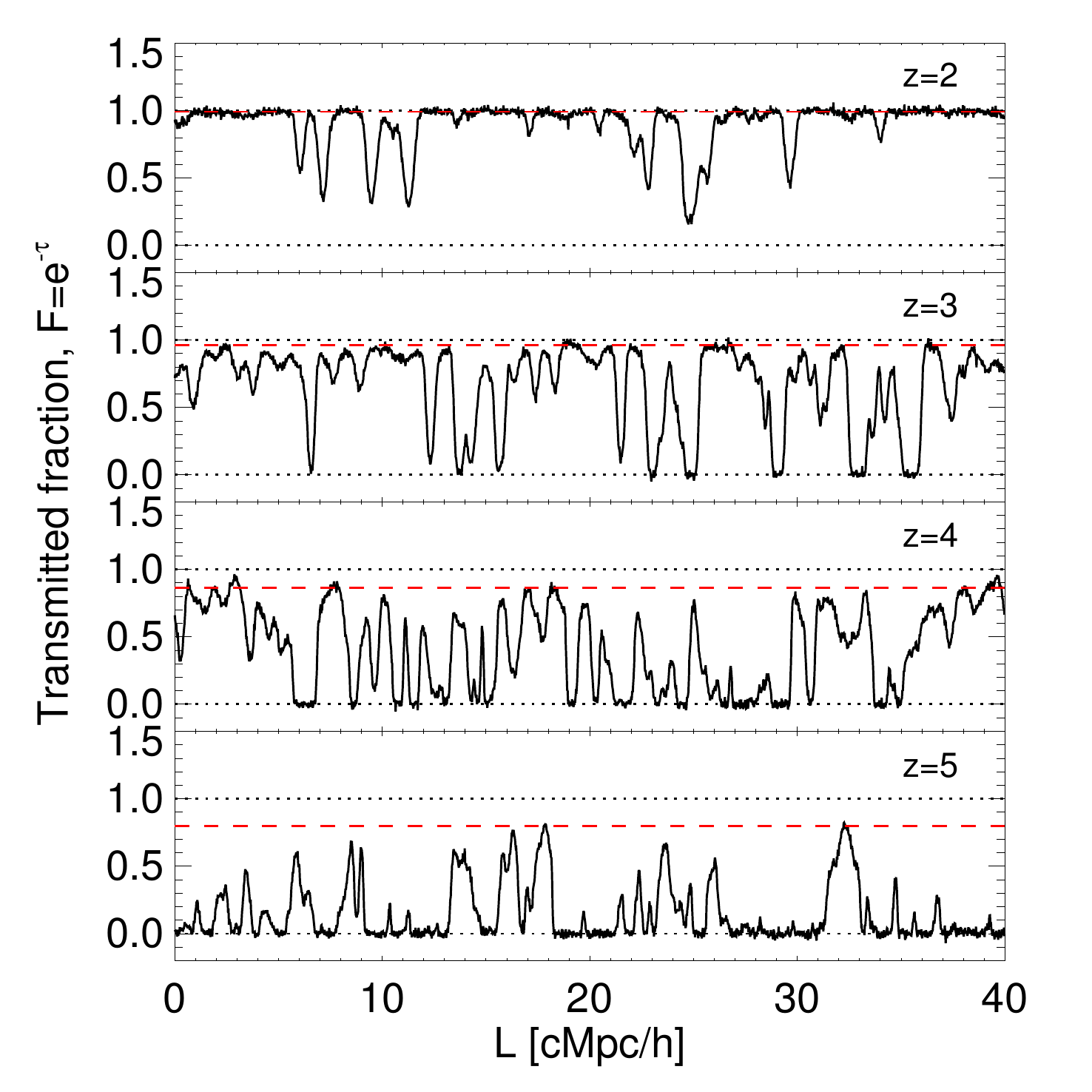}
  \vspace{-0.4cm}
  \caption{Mock \Lya forest spectra drawn from the reference 40-2048
    simulation at $z=2,\,3,\,4$ and $5$.  The spectra have been
    processed to resemble high resolution observational data (see text
    for details).  The red dashed lines display the redshift dependent
    continuum correction we adopt in this work, given by
    Eq.~(\ref{eq:Ccorr}).}
  \label{fig:spectra}
\end{center}
\end{figure}

An example of the continuum correction is displayed in
Fig.~\ref{fig:spectra}.  The mock spectra recover to the true
continuum level ($F=1$) less frequently toward higher redshift as
$\tau_{\rm eff}$ increases, hence the tendency to place the continuum
too low when performing a blind normalisation.  Note, however,
Eq.~(\ref{eq:Ccorr}) is model dependent.  If there is missing physics
in the simulations, for example volumetric heating which raises the
temperature (and hence transmission) in underdense intergalactic gas
\citep{Puchwein2012,Lamberts2015}, this may reduce the continuum bias
estimated from the models.

Lastly, to estimate the expected sample variance in the observational
data, we follow \citet{Rollinde2013} and bootstrap resample the
simulated data with replacement over the observed absorption path
length. The absorption path length interval, $\Delta X$, is related to
the redshift interval by \citep{BahcallPeebles1969}

\begin{equation} \Delta X = \frac{(1+z)^{2}}{(\Omega_{\rm m}(1+z)^{3}+\Omega_{\Lambda})^{1/2}}\Delta z. \label{eq:dX} \end{equation}

\noindent
We bootstrap with replacement 1000 times over a total path length of
$10^{5}h^{-1}\rm\,cMpc$, corresponding to $5000$ sight-lines from the
smallest simulation box size considered in this work,
$20h^{-1}\rm\,cMpc$.  For convenience, we also provide the values of
$\Delta X$ and $\Delta z$ that correspond to the box size of our
reference model, 40-2048, in Table~\ref{tab:dX}.

\begin{table}
 \centering
  \caption{Equivalent values for $L=40 h^{-1}\rm \,cMpc$ -- the box
    size of our reference model -- expressed in terms of the redshift
    interval $\Delta z$, the absorption length interval $\Delta X$,
    the distance $R=L/h(1+z)$ in proper Mpc, and velocity $v_{\rm
      H}=H(z)R$ in $\rm km\, s^{-1}$.}
 \begin{tabular}{|c|c|c|c|c|}
    \hline
    \hline
    z  & $\Delta z$ & $\Delta X$ & $R$    & $v_{\rm H}$\\
       &            &            & [pMpc] & $[\rm km\, s^{-1}]$\\
\hline
    2.0 & 0.040 & 0.120 & 19.7 & 4002\\
    2.1 & 0.042 & 0.128 & 19.0 & 4053\\
    2.5 & 0.050 & 0.163 & 16.9 & 4260\\
    2.7 & 0.054 & 0.183 & 15.9 & 4364\\
    3.0 & 0.060 & 0.213 & 14.7 & 4517\\
    3.9 & 0.081 & 0.320 & 12.0 & 4961\\
    4.0 & 0.084 & 0.334 & 11.8 & 5008\\
    5.0 & 0.109 & 0.480 & 9.8  & 5466\\
    \hline
    \hline
  \end{tabular}
 \label{tab:dX}
\end{table}


\section{Results} \label{sec:results}

\subsection{The transmitted flux distribution} \label{sec:pdf}

\begin{figure*}
  \begin{minipage}{0.85\textwidth}
    \includegraphics[width=\textwidth]{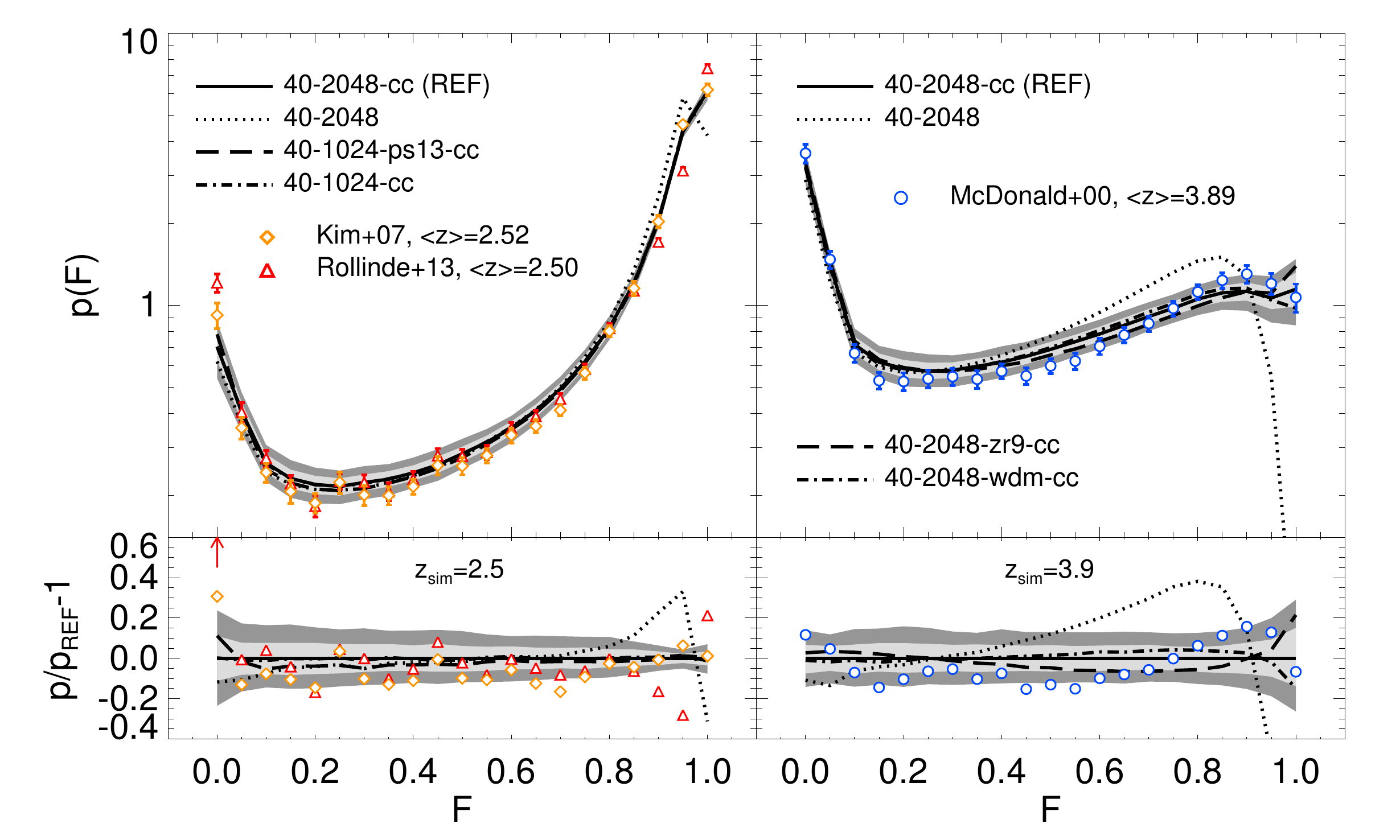}
  \end{minipage}
  \vspace{-0.2cm}
  \caption{{\it Upper panels:} The transmitted flux PDF at $z\simeq
    2.5$ (left) and $z\simeq 3.9$ (right).  The observational data are
    from \citet{Kim2007}, \citet{Rollinde2013} and
    \citet{McDonald2000}.  The solid curves correspond to the
    reference model after rescaling the effective optical depth to
    $\tau_{\rm eff}=0.227$ $(0.744)$ at $z=2.5$ $(z=3.9)$, and
    applying a redshift dependent continuum correction using
    Eq.~(\ref{eq:Ccorr}) with a small adjustment at $z=3.9$ (see text
    for details).  The grey shaded regions display the $1\sigma$ and
    $2\sigma$ uncertainties obtained when bootstrap resampling an
    absorption path length of $\Delta X=6.2$ ($z=2.5$) and $\Delta
    X=4.5$ ($z=3.9$) from a simulated path length of $10^{5}h^{-1}\rm
    cMpc$. The dotted curves in both panels show the reference model
    without the continuum correction.  The effect of star formation
    and galactic winds may be assessed by comparing the dashed and
    dot-dashed curves in the left panel.  The effect of a $3.5\rm
    \,keV$ warm dark matter particle and reionisation at $z_{\rm r}=9$
    is shown in the right panel.  {\it Lower panels:} The simulated
    and observational data (without the $1\sigma$ error bars) relative
    to the reference model. Upward pointing arrows indicate data
    points which lie outside the range of the ordinate.}
  \label{fig:pdf}
\end{figure*}

We first examine the probability distribution function (PDF) of the
transmitted flux.  This has been examined in detail in the existing
literature at $z<3$
\citep{McDonald2000,Lidz2006,Kim2007,Calura2012,Rollinde2013,Lee2015},
where discrepancies between hydrodynamical simulations and the
observational data have been highlighted
\citep{Bolton2008,Bolton2014,Tytler2009,Viel2009}.  These differences
may be due in part to missing physics in the simulations;
\citet{Bolton2008} noted the PDF measured by \citet{Kim2007} is in
better agreement with models where underdense gas is hotter than
usually expected.  Rorai et al. (2016, submitted) have recently
obtained a similar result in an analysis of the ultra-high
signal-to-noise spectrum ($\rm S/N\simeq 280$ per pixel) of quasar
HE0940$-$1050. A correction for continuum bias \citep{Lee2012} and
underestimated sample variance \citep{Rollinde2013} can also assist
with alleviating the tension between simulations and the observed PDF.

In the left panel of Fig.~\ref{fig:pdf} we compare the PDF from the
simulations to observational measurements from \citet{Kim2007} and
\citet{Rollinde2013} at $z\simeq 2.5$.  Both observational data sets
consist of 18 high resolution spectra obtained with the \emph{Very
  Large Telescope} (VLT) using the \emph{Ultraviolet and Visual
  Echelle Spectrograph} (UVES) \citep{Dekker2000}. A total of 14
spectra are common to both compilations, but the spectra have been
reduced independently.  The data in the \citet{Kim2007} redshift bin
displayed in Fig.~\ref{fig:pdf} spans $2.37 \leq z \leq 2.71$ and has
a total absorption path length of $\Delta X=6.2$.
\citet{Rollinde2013} use the same redshift bin with $\Delta X=5.3$.
The mock spectra have been scaled to match the effective optical depth
from \citet{Kim2007}, $\tau_{\rm eff}=0.227$.  The spectra are then
convolved with a Gaussian instrument profile, rebinned and noise is
added as described in Section~\ref{sec:spectra}.

The majority of the observational data at $z=2.5$ are within the
$2\sigma$ range estimated from the simulations, although around half
of the bins lie more than $1\sigma$ below the model average \citep[see
  also fig. 6 in][]{Rollinde2013}.  An isothermal ($\gamma-1 = 0$)
temperature-density relation improves the agreement at $0.1<F<0.8$
\citep[cf. fig. 4 in][]{Bolton2014}, suggesting that hot underdense
gas may be missing from the simulations.  The most significant
discrepancies, however, are at $F>0.8$ and $F=0$.  As noted by
\citet{Lee2012} the continuum correction helps improve agreement with
the \citet{Kim2007} data at $F>0.8$ -- the reference model with no continuum
correction is shown by the dotted curve.  Significant differences with
the \citet{Rollinde2013} measurements at $F \geq 0.85$ remain,
although we find a further 3 per cent change to the continuum
correction in Eq.~(\ref{eq:Ccorr}) improves the agreement.

In contrast, at $F=0$ the star formation and galactic winds
implementation and the signal-to-noise ratio are important.  This is
evident from the effect of the \citet{PuchweinSpringel2013} outflow
model\footnote{The model with the \citet{PuchweinSpringel2013} star
  formation and variable winds implementation, 40-1024-ps13, was
  performed at a lower mass resolution compared to our reference
  model.  Throughout this paper it should be compared directly to the
  40-1024 model.}, shown by the dashed curve, which increases the
number of saturated pixels by 10 per cent.  This additional absorption
is associated with saturated \Lya absorption lines with $N_{\rm
  HI}>10^{14.5}\rm\,cm^{-2}$.  We demonstrate later that these systems
are more common\footnote{We emphasise that our reference runs do not
  include star formation; cold, dense gas is instead converted into
  collisionless particles.  Differences between the 40-1024 and
  40-1024-ps13 models are therefore not only due to the influence of
  galactic outflows.} in the model with star formation and outflows
(see Section~\ref{sec:cddf}).  As pointed out by \citet{Kim2007}, the
signal-to-noise of the data also impacts on the PDF quite strongly at
$F<0.1$.  Differences in the PDF can be up to a factor of two between
$\rm S/N=25$ and $\rm S/N=100$ (see their fig. 7), with higher $\rm
S/N$ increasing the PDF at $F=0$ and decreasing it at $F=0.1$.  A more
detailed treatment of the noise may therefore also assist with
improving agreement here.

The right panel of Fig.~\ref{fig:pdf} compares the reference model to
measurements from \citet{McDonald2000} at higher redshift, $z=3.9$.
The \citet{McDonald2000} measurements are derived from eight high
resolution spectra obtained with the \emph{High Resolution Echelle
  Spectrometer} (HIRES) on the \emph{Keck} telescope \citep{Vogt1994}. The
total path length for the \citet{McDonald2000} data is $\Delta X=4.5$
and their redshift bin spans $3.39 \leq z \leq 4.43$.  The mock
spectra have been scaled to match the \citet{McDonald2000} effective
optical depth, $\tau_{\rm eff}=0.744$, and processed as before to
resemble the data.  

Without any continuum correction the reference model (dotted curve)
fails to reproduce the shape of the PDF.  We correct for this using
Eq.~(\ref{eq:Ccorr}) with a small modification, where $C_{\rm corr}$
is adjusted upwards by $0.02$ to $C_{\rm corr}=0.9$. This
significantly improves the agreement (solid curve), decreasing the
number of pixels by $30$--$40$ per cent at $F=0.8$.  Note, however,
that at $z=4$ the simulations will still underpredict the number of
pixels close to the continuum by $\sim 10$ per cent due to poor mass
resolution in the low density regions probed by the \Lya forest at
high redshift (see the convergence tests in the Appendix).  A similar
convergence rate was noted by \citet{Lukic2015} using the \textsc{Nyx}
code, suggesting that correcting for box size effects would require a
value of $C_{\rm corr}$ closer to the value predicted by
Eq.~(\ref{eq:Ccorr}) to achieve a similar level of agreement.
However, given the relatively small path length of the
\citet{McDonald2000} data, updated measurements of the PDF at $z>3.5$
would be very useful for testing this further.

\begin{figure}
\begin{center}
  \includegraphics[width=0.47\textwidth]{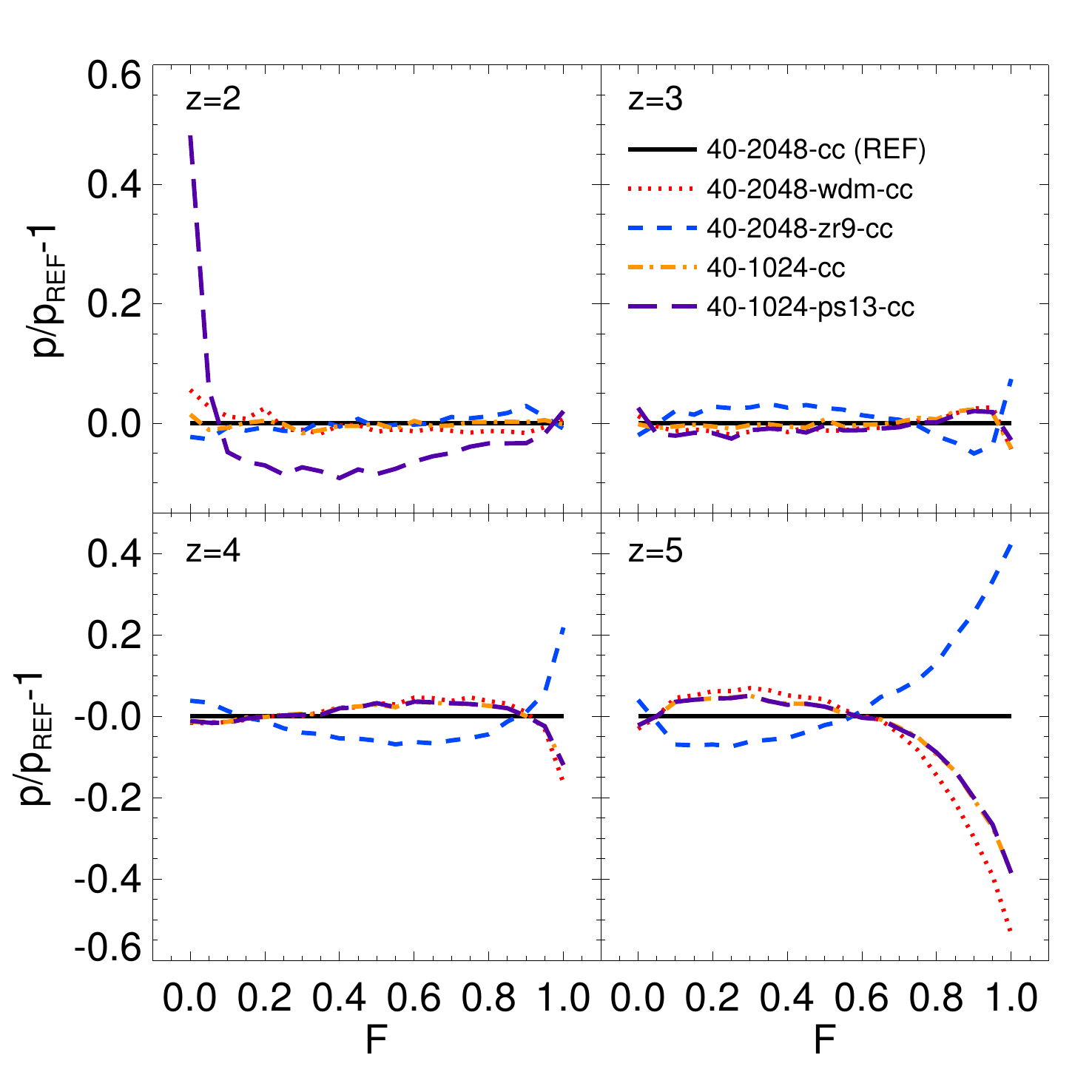}
  \vspace{-0.5cm}
  \caption{The transmitted flux PDF at $z=2,\,3,\,4$ and $5$ relative
    to the reference model (40-2048-cc).  The effect of different
    model parameters on the PDF is displayed, including a $3.5
    \rm\,keV$ warm dark matter particle (red dotted curve), rapid
    reionisation at $z_{\rm r}=9$ (blue short dashed curve) and star
    formation and galactic winds (purple long dashed curve).  The
    latter model is at lower resolution, and should be directly
    compared to the 40-1024-cc model (orange dot-dashed curve).  The
    effective optical depth is scaled to match Eq.~(\ref{eq:taueff}).}
  \label{fig:pdf_phys}
\end{center}
\end{figure}

Finally, the dashed and dot-dashed curves in the right panel of
Fig.~\ref{fig:pdf} demonstrate models with later reionisation at
$z_{\rm r}=9$ (40-2048-zr9-cc) or a $3.5\rm \, keV$ warm dark matter
particle (40-2048-wdm-cc) impact on the PDF at the 5--10 per cent
level at $z=3.9$.  These differences are smaller than the effect of a
plausible continuum correction, and largely disappear by $z<3$.  On
the other hand, star formation and galactic outflows have very little
effect on the low density gas probed by the \Lya forest at $z\simeq 4$
\citep[see also][]{Theuns2002,Viel2013feedback}.  This may be observed
more clearly in Fig.~\ref{fig:pdf_phys}, which displays the ratio of
the PDF to the reference model for a variety of model parameters at
$z=2,\,3,\,4$ and $5$ after rescaling all models to have the same
effective optical depth.  In contrast, the effect of warm dark matter
and pressure smoothing become increasingly important approaching high
redshift but act in opposite directions, decreasing and increasing the
fraction of high transmission regions, respectively.

\subsection{The transmitted flux power spectrum} \label{sec:pk}

\begin{figure*}
  \begin{minipage}{0.85\textwidth}
    \includegraphics[width=\textwidth]{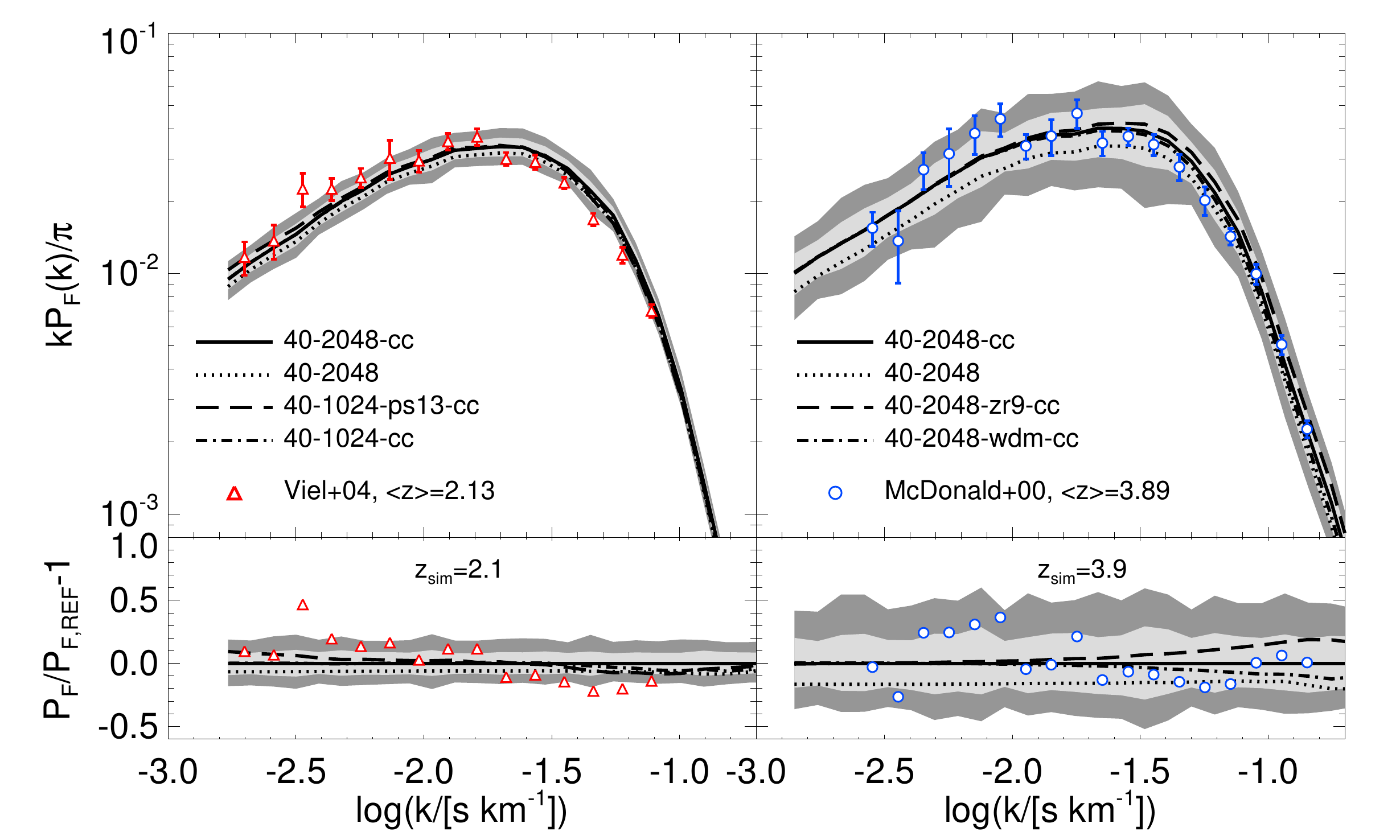}
  \end{minipage}
  \vspace{-0.2cm}
  \caption{{\it Upper panels:} The power spectrum of the transmitted
    flux estimator $\delta_{\rm F}=F/\langle F \rangle - 1$ at
    $z\simeq 2.1$ (left) and of the transmitted flux $F$ at $z\simeq
    3.9$ (right), compared to observational data from \citet{Viel2004}
    and \citet{McDonald2000}.  The solid curves correspond to the
    40-2048 model after rescaling the effective optical depth to
    $\tau_{\rm eff}=0.180$ $(0.744)$ at $z=2.1$ $(z=3.9)$, and
    applying a redshift dependent continuum correction using
    Eq.~(\ref{eq:Ccorr}) with a small adjustment at $z=3.9$ (see text
    for details).  The grey shaded regions display the 68 and 95 per
    cent confidence intervals obtained when bootstrap resampling an
    absorption path length of $\Delta X=11.9$ ($\Delta X=4.5$) from a
    simulated path length of $10^{5}h^{-1}\rm cMpc$.  The other curves
    are as described in Fig.~\ref{fig:pdf}. {\it Lower panels:} The
    simulations and observational data (excluding error bars) relative
    to the reference model 40-2048-cc.}
      \label{fig:pk}
\end{figure*}

The power spectrum of the transmitted flux has been extensively used
to constrain cosmological parameters with the \Lya forest
\citep{Croft1999,Croft2002,
  McDonald2000,McDonald2006,Viel2004,Viel2013,Palanque2013,Palanque2015}. A
comparison of the simulations to measurements of the power spectrum at
$z=2.1$ and $z=3.9$ is displayed in Fig.~\ref{fig:pk}.

The left panel of Fig.~\ref{fig:pk} compares the measurements of
\citet{Viel2004} at $\langle z \rangle=2.13$ to the models.  The
\citet{Viel2004} data are obtained from 16 quasar spectra spanning the
redshift bin $2<z<2.3$, selected from the \emph{VLT/UVES} LUQAS (Large
Sample of UVES Quasar Absorption Spectra) sample from \citet{Kim2004}.
The \citet{Viel2004} power spectrum is computed using the estimator
$\delta_{\rm F}=F/\langle F \rangle - 1$.  The absorption path length
of these data is $\Delta X=11.9$ and our mock spectra are rescaled to
correspond to $\tau_{\rm eff}=0.18$, consistent with the value of
$\tau_{\rm eff}=0.17\pm 0.02$ quoted by \citet{Viel2004}.  These
authors inferred cosmological parameters by fitting the power spectrum
in the range $-2.5 < \log(k/\rm s\,km^{-1}) < -1.5$, where systematic
uncertainties due to continuum fitting, damped absorption systems and
metal lines should be minimised.

We find the Sherwood simulations are in good agreement with the data,
with the majority of the bins lying within $1$--$2\sigma$ of the
models.  The largest discrepancy is the data point at $\log (k/\rm
s\,km^{-1}) \simeq -2.5$, although the dashed curve corresponding to
40-1024-ps13-cc again indicates that star formation and outflows may
improve agreement here.  The larger number of high column density
systems in this model increases the power by around 10 per cent at
large scales, $\log (k/\rm s\,km^{-1}) < -2$ \citep[see
  also][]{Viel2004strong,McDonald2005}.  In contrast, the impact of
the \citet{PuchweinSpringel2013} star formation and winds at scales
$\log(k/\rm s\,km^{-1})>-1.5$ is rather small; note that much of the
difference relative to the reference run is due to the lower mass
resolution of the 40-1024-ps13-cc model.  Instead, hotter underdense
gas can decrease power on small scales at this redshift, potentially
improving agreement with the small scale measurements \citep[see
  e.g. fig. 3 in][]{Viel2004}.  The continuum correction is small at
$z=2.1$, but acts to increase power on all scales by $\sim 5$ per
cent.

A higher redshift measurement at $\langle z \rangle=3.89$ from
\citet{McDonald2000} is compared to the models in the right hand panel
of Fig.~\ref{fig:pk}. The mock spectra have been scaled to $\tau_{\rm
  eff}=0.744$, and the observational data, absorption path length and
continuum correction are the same as those used for the
\citet{McDonald2000} PDF measurement at the same redshift (see
Section~\ref{sec:pdf}).  The reference model is in excellent agreement
($1$--$2\sigma$) with data following the continuum correction, which
increases power on all scales.  Without this correction, however, the
data points lie systematically above the model, particularly at scales
$\log(k/\rm s\,km^{-1}) < -1.5$.  This suggests a careful treatment of
the continuum placement is especially important for forward modelling
high redshift \Lya forest data.  Although the effect of star formation
and galactic outflows is minimal at $z\simeq 4$ (see
Fig.~\ref{fig:pk_phys}), a later reionisation at $z_{\rm r}=9$ results
in less pressure smoothing and hence more power on scales $\log(k/\rm
s\,km^{-1}) > -1.7$.  This indicates it should be possible to
constrain the integrated thermal history during reionisation using the
line of sight \Lya forest power spectrum at high redshift
\citep{Nasir2016}.  However, the magnitude of the effect at $z \simeq
4$ lies within the $1\sigma$ bootstrapped uncertainty for the
\citet{McDonald2000} absorption path length.  Either more quasar
spectra or higher redshift data where the effect is stronger will
assist here.  Note also the opposite effect is observed for the warm
dark matter model, with power reduced by around 10 per cent at scales
$\log(k/\rm s\,km^{-1}) \sim -1$.  Again, however, this difference
lies within the expected $1\sigma$ range for the rather small
absorption path length of the \citet{McDonald2000} sample.

\begin{figure}
\begin{center}
  \includegraphics[width=0.47\textwidth]{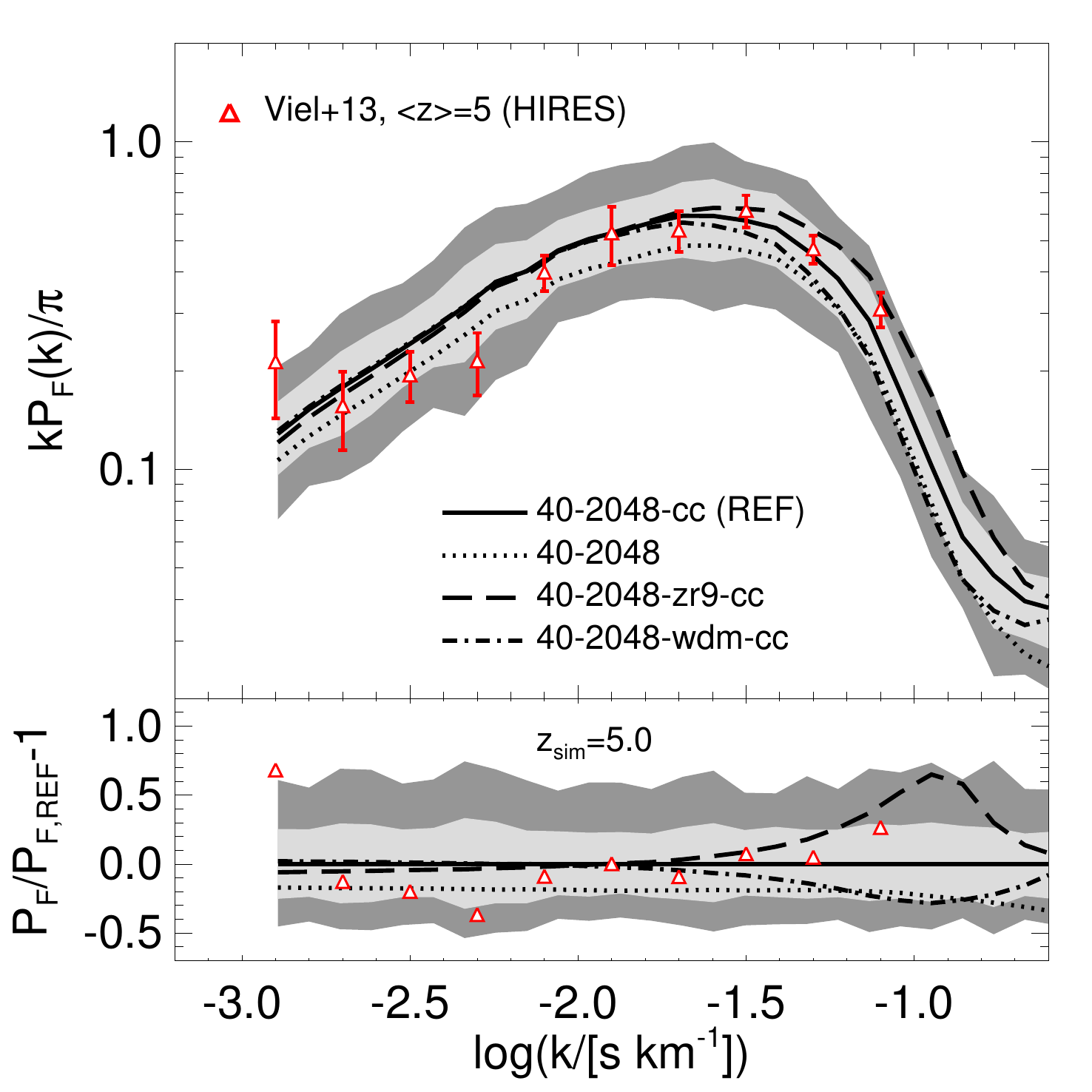}
  \vspace{-0.3cm}
  \caption{{\it Upper panel:} The power spectrum of the transmitted
    flux estimator $\delta_{\rm F}=F/\langle F \rangle - 1$ at $z=5$,
    compared to observational data from \citet{Viel2013} obtained with
    the \emph{Keck/HIRES} spectrograph.  The solid curve corresponds
    to the 40-2048-cc model with $\tau_{\rm eff}=1.76$.  The grey
    shaded regions display the $1\sigma$ and $2\sigma$ uncertainties
    obtained when bootstrap resampling an absorption path length
    interval of $\Delta X=5.4$ from a simulated path length of
    $10^{5}h^{-1}\rm cMpc$. The dotted curve shows the 40-2048 model
    without the continuum correction.  The effect of a $3.5\rm\, keV$
    warm dark matter particle and rapid reionisation at
    $z_{\rm r}=9$ are shown by the dashed and dot-dashed curves. The
    mock spectra have been processed to match the resolution and
    signal-to-noise properties of the observational data (see text for
    details). {\it Lower panel:} The simulated and observational data
    (excluding error bars) relative to the reference model.}
  \label{fig:pk_viel}
\end{center}
\end{figure}

In Fig.~\ref{fig:pk_viel} we also compare the simulations to the power
spectrum at $\langle z \rangle=5$ presented by \citet{Viel2013}.
These authors placed a lower limit on the possible mass of a warm dark
matter thermal relic, $m_{\rm WDM} \geq 3.3\rm\,keV$ ($2\sigma$).  The
power spectrum measurement is based on an analysis of 14 quasar
spectra with emission redshifts $4.48 \leq z \leq 6.42$ obtained with
\emph{Keck/HIRES}.  The redshift bin displayed here spans the redshift
range $4.8<z<5.2$.  The total path absorption path length is $\Delta
X=5.4$ with a typical $\rm S/N = 10$--$20$.  The simulations have been
rescaled to $\tau_{\rm eff}=1.76$, consistent with the lower $1\sigma$
bound on the best fit obtained by \citet{Viel2013}, and a uniform
signal-to-noise of $\rm S/N=20$ is added to each pixel.  Note that to
obtain their warm dark matter constraints, \citet{Viel2013} use the
power spectrum over the range $-2.3 \leq \log(k/\rm s\,km^{-1}) \leq
-1.1$.

As with the comparison at $z\simeq 3.9$, the continuum correction
increases the power by around 20 per cent on all scales.  The
observations are again in very good agreement with the reference
model; most lie within $1$--$2\sigma$ of the expected variance.  It is
also clear that the differences between the reference and late
reionisation models are more apparent toward higher redshift, closer
to the reionisation redshift and where the \Lya forest probes scales where
structure is closer to the linear regime.  Note, however, the mass
resolution correction means the power on scales $\log (k/\rm
s\,km^{-1}) \simeq -1$ is still underestimated by around 10 per cent
in this model \citep[see the Appendix
  and][]{BoltonBecker2009,Viel2013}.

Finally, Fig.~\ref{fig:pk_phys} displays the power spectrum for
different model parameters at four different redshifts, where again
all models are scaled to match Eq.~(\ref{eq:taueff}).  As for the PDF,
the effect of star formation and galactic winds is significant at
$z<3$, increasing the power on large scales due to the presence of
additional high column density systems (see Section~\ref{sec:cddf}).
Similarly, the effect of a $3.5\rm\, keV$ warm dark matter thermal
relic and late reionisation are degenerate and have a dramatic
effect on the power on small scales at high redshift (40--80 per
cent), but only at the level of a few per cent at lower redshift.

\begin{figure}
\begin{center}
  \includegraphics[width=0.47\textwidth]{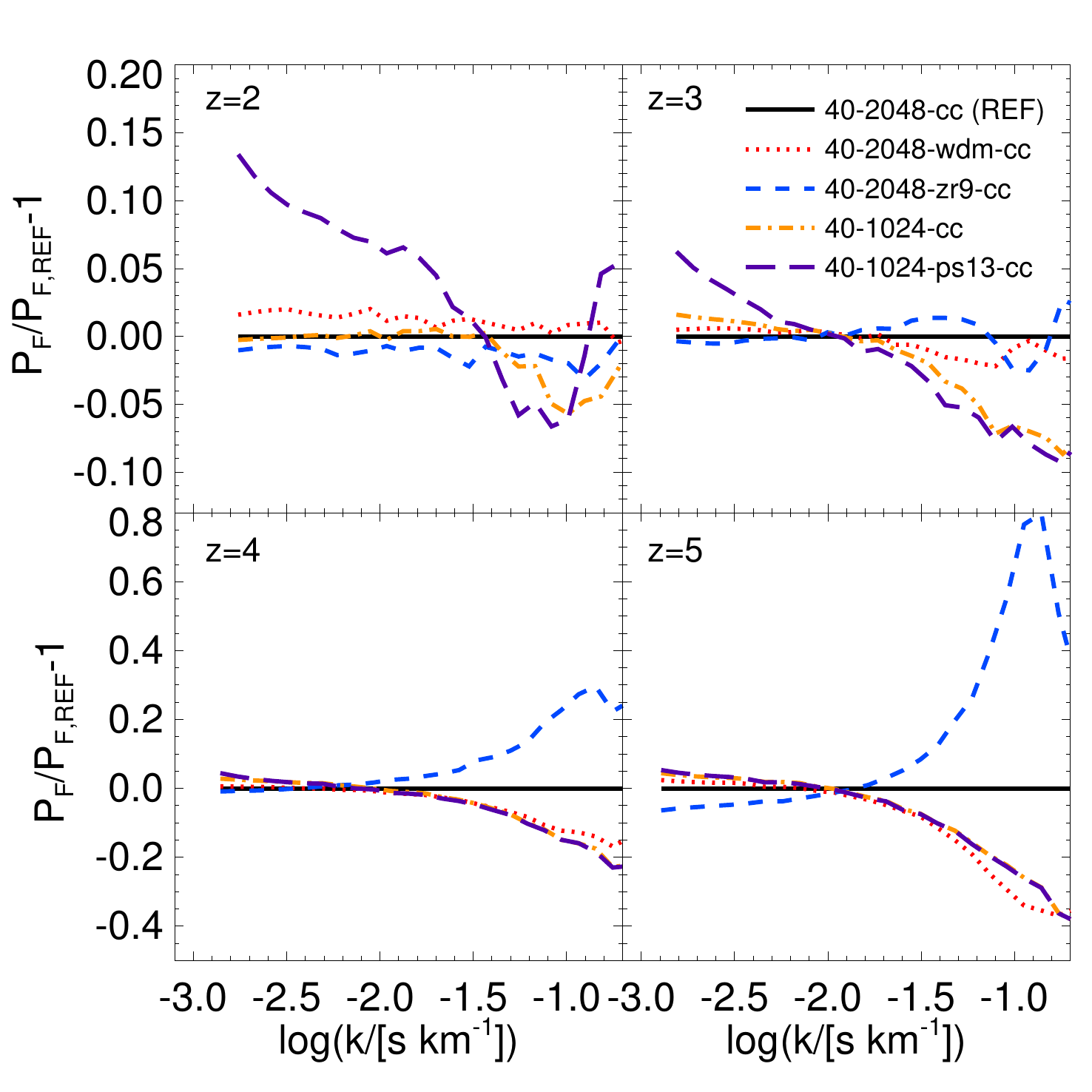}
  \vspace{-0.4cm}
  \caption{The power spectrum of the transmitted flux, $F$, at
    $z=2,\,3,\,4$ and $5$ relative to the reference model
    (40-2048-cc).  The effect of different model parameters on the
    power spectrum is displayed, including a $3.5\rm\,keV$ warm dark
    matter particle (red dotted curve), rapid reionisation at
    $z_{r}=9$ (blue short dashed curve) and star formation and
    outflows (purple long dashed curve).  The latter model is at lower
    resolution, and should be directly compared to the 40-1024-cc model
    (orange dot-dashed curve).  The effective optical depth is scaled
    to match Eq.~(\ref{eq:taueff}).}
  \label{fig:pk_phys}
\end{center}
\end{figure}

\subsection{The standard deviation of the mean transmitted flux}

\begin{figure}
\begin{center}
  \includegraphics[width=0.47\textwidth]{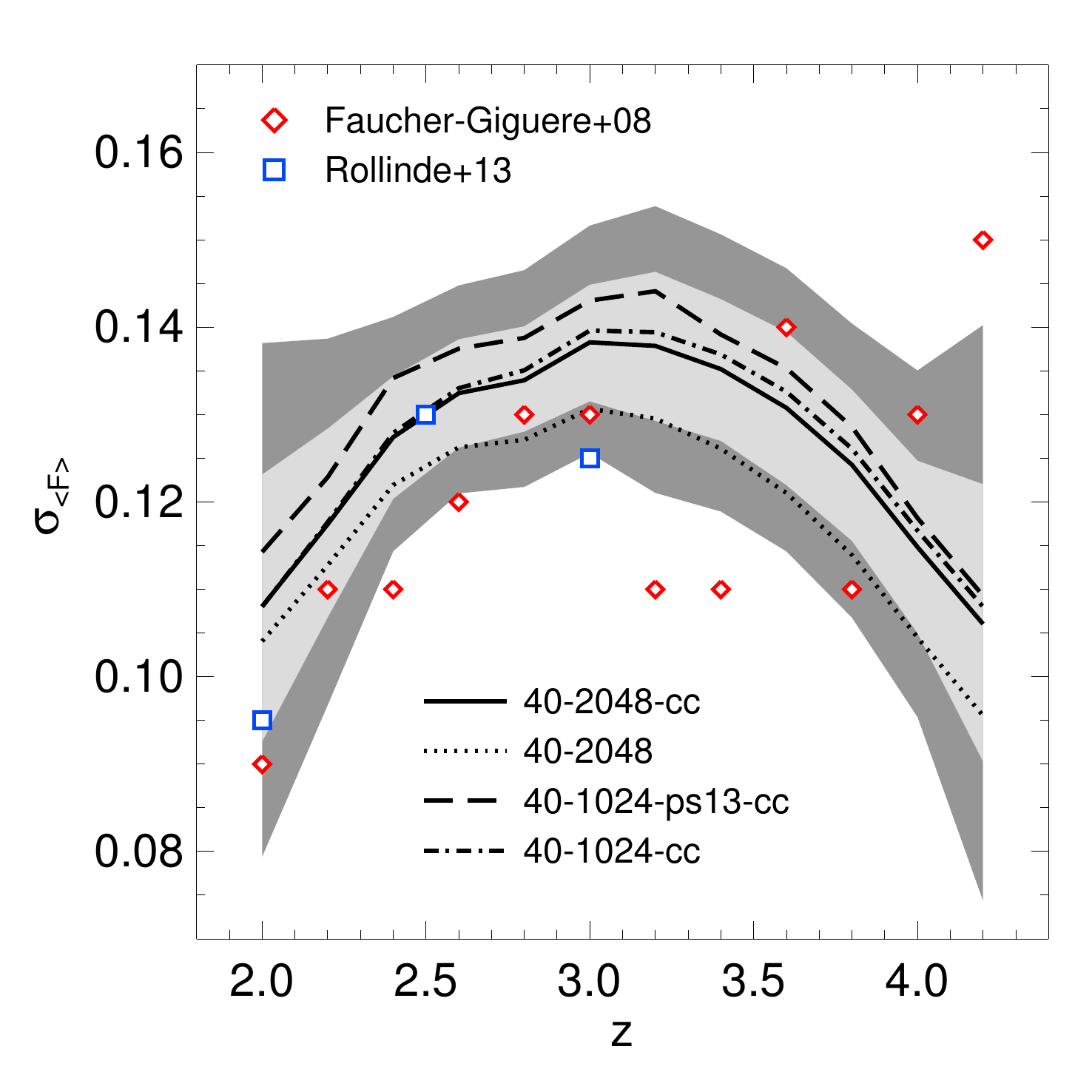}
  \vspace{-0.5cm}
  \caption{The standard deviation of the mean transmitted flux
    measured in $3\rm \,pMpc$ segments as a function of redshift.  The
    red diamonds and blue squares are observational measurements from
    \citet{FaucherGiguere2008} and \citet{Rollinde2013}. Simulation
    results are represented by the curves indicated in the panel.
    All simulation results, aside from the dotted curve, also have a
    redshift dependent continuum correction applied (see
    Eq.~\ref{eq:Ccorr}).  The grey shaded regions correspond to the 68
    and 95 per cent confidence intervals for the 40-2048 model,
    obtained when bootstrap resampling the \citet{FaucherGiguere2008}
    path length in each redshift bin from a total simulated path
    length of $10^{5}h^{-1}\rm\,cMpc$.}
  \label{fig:sigmaF}
\end{center}
\end{figure}

We next investigate the scatter in the mean transmission by comparing
the standard deviation of the mean transmitted flux, $\sigma_{\langle
  F \rangle}$, to observational measurements from
\citet{FaucherGiguere2008} and \citet{Rollinde2013} (the latter data
set is described in Section~\ref{sec:pdf}).  The
\citet{FaucherGiguere2008} data consist of 86 high resolution quasar
spectra obtained with the \emph{Echelle Spectrograph and Imager} (ESI)
and HIRES on the \emph{Keck} telescope and the \emph{Magellan Inamori Kyocera
  Echelle} (MIKE) spectrograph on the \emph{Magellan} telescope.  Our mock
spectra are processed to match the best-fit power law to the
effective optical depth from \citet{FaucherGiguere2008}, $\tau_{\rm
  eff}=0.0018(1+z)^{3.92}$.  Following these authors, at intervals of
$\Delta z=0.2$ over the range $2\leq z \leq 4.2$ we compute the
standard deviation of the mean transmitted flux, $\sigma_{\langle F
  \rangle}$, in segments of $3\rm\,pMpc$.  We estimate the sample
variance on this quantity by bootstrap resampling the observed path
length from \citet{FaucherGiguere2008} (see their fig. 4) from a total
path length of $10^{5}h^{-1}\rm\,cMpc$.

Fig.~\ref{fig:sigmaF} shows the standard deviation peaks around $z=3$,
and falls toward lower and higher redshift as proportionally more
pixels lie either at the continuum ($F=1$) or are saturated ($F=0$).
The 68 and 95 per cent confidence intervals for the continuum
corrected reference model are indicated by the grey shaded regions.
The simulations are in agreement with the data at $z<3$, but at $z>3$
three data points differ at more than $2\sigma$.  It is unlikely these
differences are attributable to star formation and galactic
outflows. A comparison of the dot-dashed and dashed curves in
Fig.~\ref{fig:sigmaF} demonstrate the \citet{PuchweinSpringel2013}
variable winds model increases $\sigma_{\langle F \rangle}$ by $\sim
8$ per cent at $z=2$, but this difference diminishes toward higher
redshift.  In contrast, the impact of the continuum correction --
indicated by the solid and dotted curves -- is much larger.  This acts
to increase $\sigma_{\langle F \rangle}$ by distributing the pixels
over a broader range of values.  Lowering the continuum at $z=4.2$ by
a further $10$ per cent, or placing the continuum too high by $5$ per
cent at $z=3.2$ and $z=3.4$, leads to $2\sigma$ agreement with the
data.  This suggests that observational systematics may in part
explain the discrepancy at $z>3$.  We have verified the reference
simulation is converged to within $2$--$3$ per cent with respect to
box size and mass resolution over the redshift range we consider;
further details are available in the Appendix.

\subsection{The column density distribution function} \label{sec:cddf}

In this section we now turn to compare our mock spectra to the results
of a Voigt profile analysis of high resolution data using
VPFIT\footnote{\url{http://www.ast.cam.ac.uk/~rfc/vpfit.html}}
\citep{CarswellWebb2014}.  The observational data are from
\citet{Kim2013}, which comprises 18 high resolution quasar spectra
from the \emph{European Southern Observatory} (ESO) \emph{VLT/UVES}
archive with a typical signal-to-noise per pixel of $\rm
S/N=35$--$50$.   We compare to the \Lya forest observed in
two redshift bins at $\langle z \rangle =2.13$ and $\langle z \rangle
=2.72$, spanning $1.9 < z < 2.4$ and $2.4 < z < 3.2$ with a total
absorption path length of $\Delta X=12.5$ and $\Delta X=10.5$,
respectively.  Absorption features within $\pm 50\rm\, \AA$ of damped
\Lya absorbers are ignored in these data, as well as all \Lya lines
within $5000\rm \, km\,s^{-1}$ of the quasar systemic redshift to
avoid the proximity effect.  We use the \HI column densities, $N_{\rm
  HI}$, and velocity widths, $b_{\rm HI}$, obtained from fits to the
\Lya absorption profiles only, enabling a straightforward comparison
to the mock \Lya forest spectra.  We note, however, that both
\citet{Kim2013} and \citet{Rudie2013} provide alternative fits for
saturated lines, $N_{\rm HI}>10^{14.5}\rm\,cm^{-2}$, obtained using
higher order Lyman series transitions.

The mock spectra are rescaled to match the effective optical depth
evolution from \citet{Kim2007}, $\tau_{\rm eff}=0.143$ $(0.273)$ at
$z=2.1$ ($z=2.7$), and processed as described in
Section~\ref{sec:spectra}.  Voigt profiles were then fitted to
$20h^{-1}\rm \, cMpc$ segments using VPFIT.  All lines within
$100\rm\,km\,s^{-1}$ of the start and end of each segment were ignored
to avoid edge artefacts. We use the same procedure followed by
\citet{Kim2013} when performing this analysis, thus minimising any
possible biases introduced by the Voigt profile decomposition process.

We first examine the column density distribution function (CDDF),
defined as the number of absorption lines per unit column density per
unit absorption path length

\begin{equation} f(N_{\rm HI},X) = \frac{\partial^{2}n}{\partial N_{\rm HI}\partial X}. \end{equation}

\noindent
We compare to the \citet{Kim2013} CDDF measurements for absorption
lines in the \Lya forest with column densities in the range
$10^{12.3}\rm\, cm^{-2} \leq N_{\rm HI} \leq 10^{16.8}\rm \,cm^{-2}$.
Note, however, that at $N_{\rm HI}<10^{12.7}\rm\,cm^{-2}$ the CDDF
will be affected by incompleteness.  Absorption lines that are
optically thick to Lyman continuum photons (i.e $N_{\rm
  HI}>10^{17.2}\rm\,cm^{-2}$) will not be captured in our optically
thin simulations, and we do not consider these further here.  Recent
hydrodynamical simulations that incorporate corrections for
self-shielding have been shown to capture the normalisation and shape
of the CDDF in this regime \citep{Altay2011,Rahmati2013}.

\begin{figure*}
  \begin{minipage}{0.85\textwidth}
    \includegraphics[width=\textwidth]{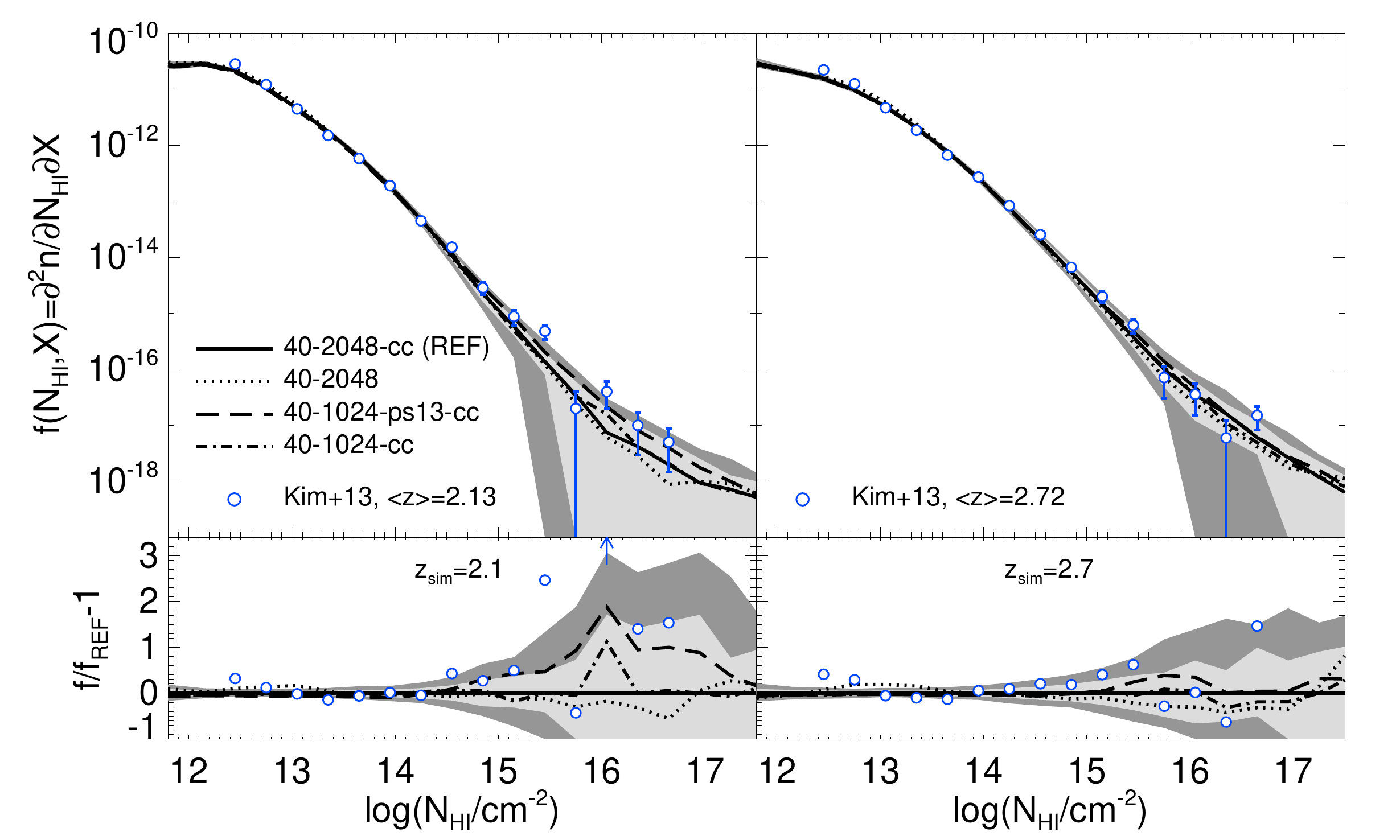}
  \end{minipage}
  \vspace{-0.2cm}
  \caption{{\it Upper panels:} The \Lya forest column density
    distribution function compared to observational data from
    \citet{Kim2013} at $z\simeq 2.1$ (left) and $z \simeq 2.7$
    (right).  The solid curves correspond to the 40-2048-cc model
    after rescaling the effective optical depth to match
    \citet{Kim2007}, $\tau_{\rm eff}=0.143$ ($0.273$) at $z=2.1$
    ($z=2.7$), and applying the redshift dependent continuum
    correction from Eq.~(\ref{eq:Ccorr}).  The grey shaded regions
    display the 68 and 95 per cent confidence intervals obtained when
    bootstrap resampling an absorption path length interval of $\Delta
    X=12.5$ ($z=2.1$) and $\Delta X=10.5$ ($z=2.7$) from a simulated
    path length of $10^{5}h^{-1}\rm cMpc$. The dotted curve shows the
    40-2048 model without the continuum correction.  The effect of
    galactic winds may be assessed by comparing the dashed and
    dot-dashed curves.  The mock spectra have been processed to
    broadly match the resolution and signal-to-noise properties of the
    observational data (see text for details). {\it Lower panel:} The
    simulated and observational data (excluding error bars) relative
    to the reference model.  Upward pointing arrows indicate data
    points which lie outside the range of the ordinate.}
  \label{fig:cddf}
\end{figure*}

Fig.~\ref{fig:cddf} displays the observed CDDF in comparison to
simulations at $z=2.1$ (left) and $z=2.7$ (right).  The agreement,
particularly for unsaturated lines with $10^{12.7}\rm\,cm^{-2}<N_{\rm
  HI}<10^{14.4}\rm\,cm^{-2}$ is remarkably good and is largely within
the expected $1\sigma$ and $2\sigma$ bootstrapped uncertainties
displayed by the grey shading.  The dotted curve indicates the
continuum correction acts to slightly decrease (increase) the number
of weak (strong) lines.  On the other hand, the simulations
underestimate the number of lines at $N_{\rm
  HI}=10^{12.4}\rm\,cm^{-2}$ by $30$--$40$ per cent.  The comparison
between the solid and dot-dashed curves indicates the simulations are
well converged with respect to mass resolution here (see also the
Appendix).  However, these low column density lines are strongly
affected by the signal-to-noise of the data and misidentified metal
lines.  We have verified that lowering the signal-to-noise further
decreases the number of weak lines identified.  As our simulations do
not include metals and adopt a uniform $\rm S/N=50$, this may in part
account for the difference.

\begin{figure*}
  \begin{minipage}{0.85\textwidth}
    \includegraphics[width=\textwidth]{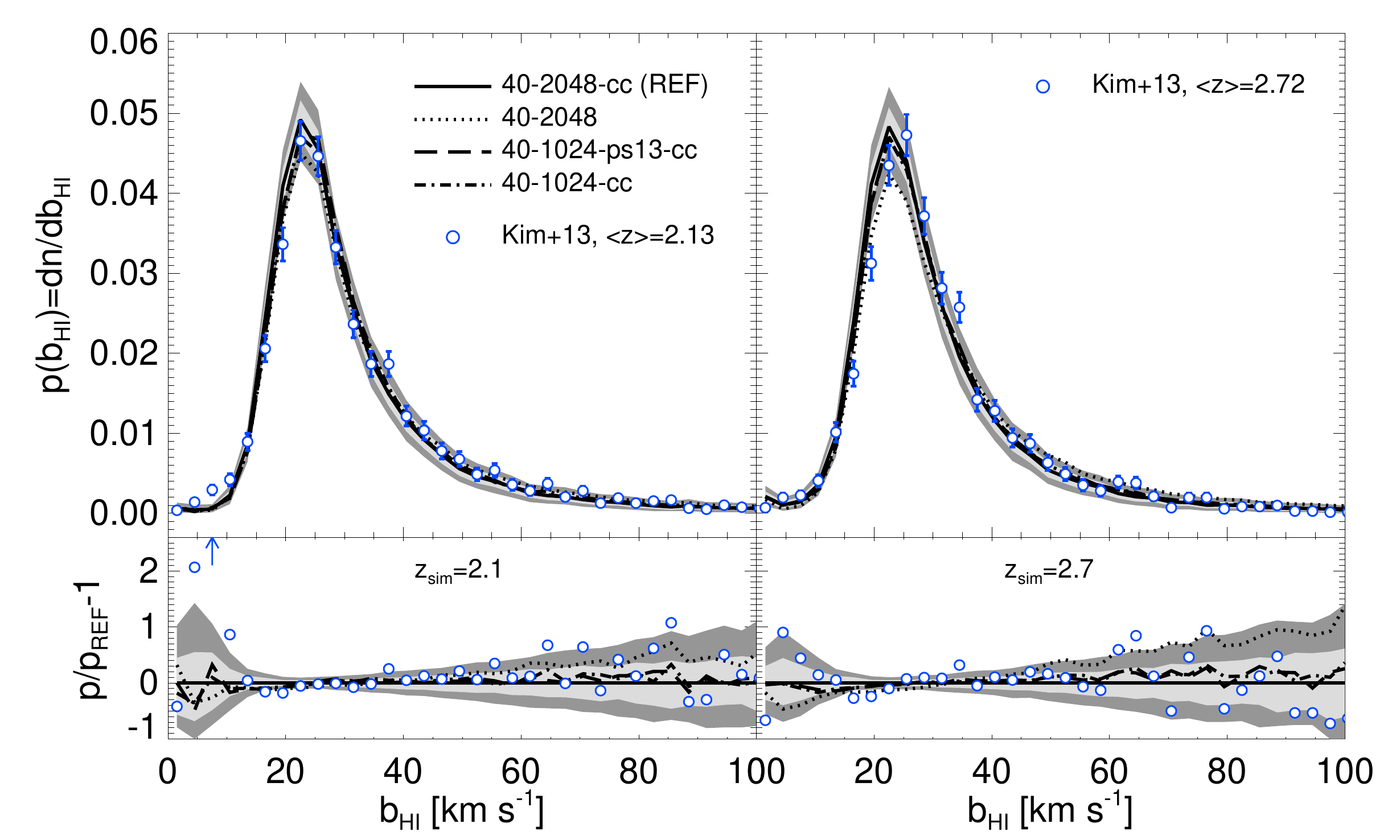}
  \end{minipage}
  \vspace{-0.2cm}
  \caption{{\it Upper panels:} The distribution of velocity widths,
    $b_{\rm HI}$, for \Lya absorption lines compared to observational
    data from \citet{Kim2013} at $z=2.1$ (left) and $z=2.7$ (right).
    Only lines with $12.3 \leq \log(N_{\rm HI}/\rm cm^{-2}) \leq 17.0$
    are included in this comparison. The simulation data, shown by the
    curves and grey shading, are as described in the caption of
    Fig.~\ref{fig:cddf}. {\it Lower panels:} The simulated and
    observational data (excluding error bars) relative to the
    reference model. Upward pointing arrows indicate data points which
    lie outside the range of the ordinate.}
  \label{fig:bdist}
\end{figure*}

For saturated lines at $N_{\rm HI} \geq 10^{14.5}\rm\,cm^{-2}$, a much
larger difference is seen for the model with the
\citet{PuchweinSpringel2013} star formation and galactic winds
implementation at $z=2.1$.  Comparing the dot-dashed and dashed curves
suggest the incidence of saturated lines is underestimated by up to a
factor of two for models which do not include sub-resolution
treatments for galaxy formation physics and galactic outflows.  This
suggests the approximate scheme for removing cold $(T<10^{5}\rm\,K$),
dense ($\Delta>1000$) gas used in our reference runs does not fully
capture the incidence of saturated absorption systems at $z\simeq 2$.
These differences are qualitatively consistent with the results from
the PDF and power spectrum, where the observational data exhibit more
saturated pixels and an increase in large scale power compared to the
simulations at similar redshifts. In contrast, at $z=2.7$ the effect
of star formation and galactic outflows on the CDDF is more modest and
the incidence of saturated lines in the mock spectra is consistent
with the observations.  This indicates that the sensitivity of the
\Lya forest to high density gas decreases toward higher redshift.

\subsection{The velocity width distribution} \label{sec:bdist}

The distribution of \Lya line velocity widths, $b_{\rm HI}$, from the
\citet{Kim2013} observational data are displayed in
Fig.~\ref{fig:bdist}.  Only the velocity widths for absorbers with
$10^{12.3}\rm\,cm^{-2}\leq N_{\rm HI}\leq 10^{17.0}\rm\,cm^{-2}$ are
included in the distribution.  The simulations are again in very good
agreement with the data.  In this case, the continuum correction acts
to produce more (fewer) narrow (wide) lines.  At $z=2.1$ there is a
deviation at just over $2\sigma$ for the bin at $20\rm\,km\,s^{-1}$,
although the peak of the simulated distribution matches the data.
However, the main discrepancy at $z=2.1$ (and to a much lesser extent
at $z=2.7$) is for lines with $b_{\rm HI}<10\rm\,km \,s^{-1}$.  In
combination with missing low column density lines in the CDDF, this
again suggests misidentified metal line absorbers -- which become more
prevalent toward lower redshift -- are responsible for the additional
weak, narrow lines in the data.  The excellent agreement with the
shape of the distribution at $b_{\rm HI}>10\rm\,km\,s^{-1}$ indicates
that the gas temperatures at this redshift are largely consistent with
the observational data.  This is expected given the models were tuned
to match (independent) observational constraints on the IGM
temperature (e.g. Fig.~\ref{fig:Trho}).

In the higher redshift bin, the agreement is also generally good.  The
peak of the velocity width distribution in the simulations is $3\rm\,
km\, s^{-1}$ below the peak in the observed distribution, but remains
consistent within the expected $2\sigma$ range.  Two data points at
$17\rm \,km\,s^{-1}$ and $20\rm \,km\,s^{-1}$ again deviate at more
than $2\sigma$, possibly indicating the gas may be slightly too cold
(by a few thousand degrees) in the model at this redshift.  Additional
heating from non-equilibrium ionisation effects during \HeII
reionisation may account for this small difference
\citep{Puchwein2015}.

\begin{figure}
\begin{center}
  \includegraphics[width=0.47\textwidth]{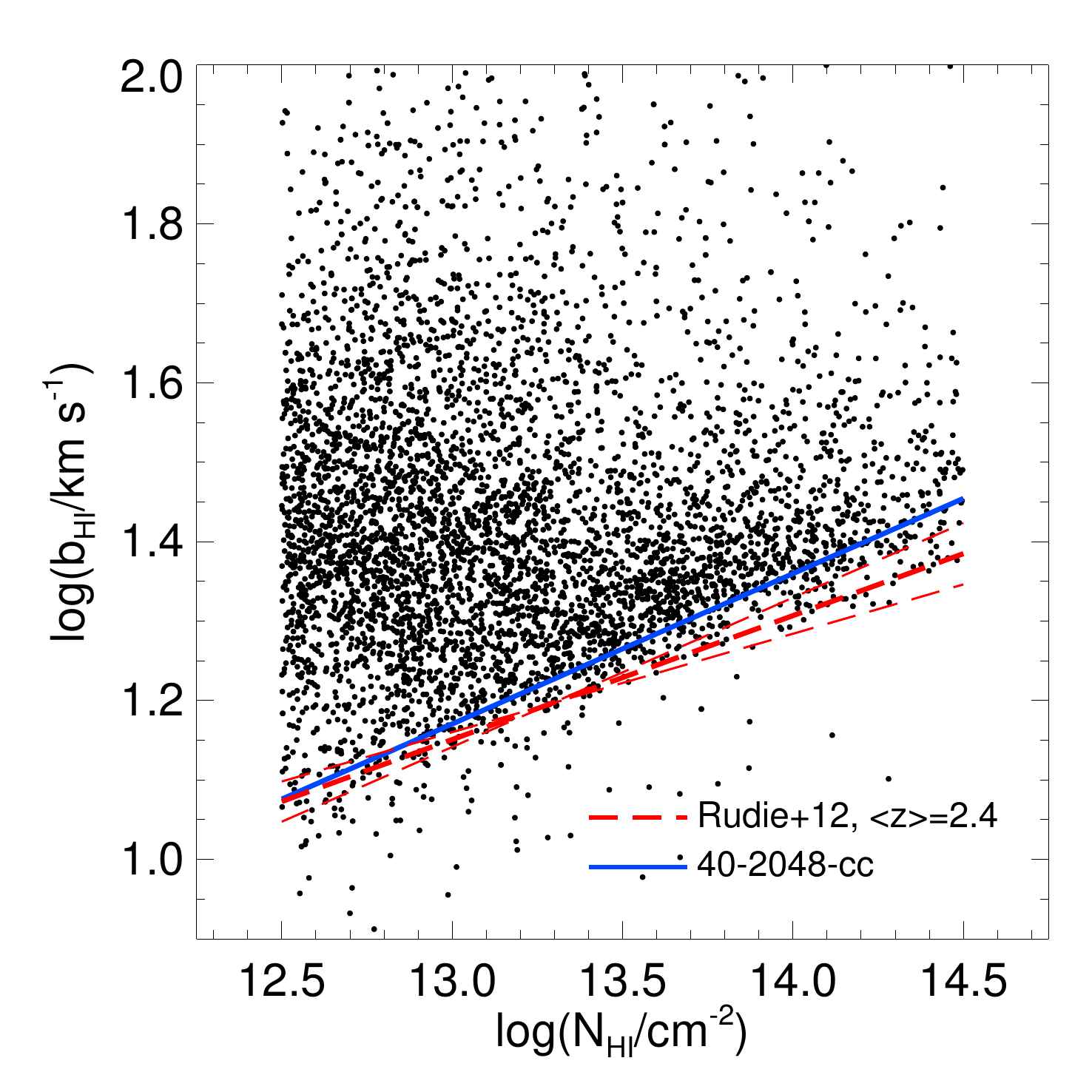}
  \vspace{-0.4cm}
  \caption{The $b_{\rm HI}-N_{\rm HI}$ plane at $z=2.4$, shown by the
    points, corresponds to $5\,000$ Voigt profiles which have been
    fitted to mock spectra from the reference model.  The effective
    optical depth is scaled to match Eq.~(\ref{eq:taueff}) and a
    redshift dependent continuum correction has been applied using
    Eq.~(\ref{eq:Ccorr}).  The signal to noise is $\rm S/N=89$, and
    the lines have been selected following the default criteria
    specified by \citet{Rudie2012}.  The lower envelope of this
    distribution may be compared to the observationally measured lower
    cut-off from \citet{Rudie2012}. The thick red dashed line
    corresponds to their ''default'' measurement, with the thin lines
    representing the $1\sigma$ uncertainty. For comparison, the blue
    line shows the lower cut-off measured in the same way from the
    simulation data.}
  \label{fig:bNplane}
\end{center}
\end{figure}

As a further consistency check, in Fig.~\ref{fig:bNplane} we examine
the $b_{\rm HI}$--$N_{\rm HI}$ plane at $z=2.4$ in the simulations.
This is compared to the lower envelope of the $b_{\rm HI}$--$N_{\rm
  HI}$ plane measured by \citet{Rudie2012} from simultaneous fits to
\Lya and \Lyb absorption in a sample of 15 high signal-to-noise ($\rm
S/N \simeq 50$--$200$) quasar spectra obtained with \emph{Keck/HIRES}.
Their Voigt profile analysis consists of $5758$ absorbers with $12.0 <
\log(N_{\rm HI}/\rm cm^{-2}) <17.2$ and a mean redshift of $\langle z
\rangle=2.37$.  \citet{Rudie2012} fit a power-law to the lower
envelope of the $b_{\rm HI}$--$N_{\rm HI}$ distribution for $12.5 <
\log(N_{\rm HI}/\rm cm^{-2}) <14.5$ following the method of
\citet{Schaye1999}.  By comparing this power-law relation to models,
it is possible to infer the amplitude and slope of the temperature
density relation, $T=T_{0}\Delta^{\gamma-1}$.  \citet{Bolton2014}
performed a detailed reanalysis of this measurement, finding
$T_{0}=10000^{+3200}_{-2100}\rm \,K$ and $\gamma-1=0.54\pm 0.11$.

We scale the effective optical depth of mock spectra at $z=2.4$ to
match Eq.~(\ref{eq:taueff}) and add a uniform signal-to-noise per
pixel of $\rm S/N=89$, corresponding to the average of the
\citet{Rudie2012} data set.  Lines are selected following the default
criteria described by these authors.  The resulting $b_{\rm
  HI}$--$N_{\rm HI}$ plane in Fig.~\ref{fig:bNplane} is broadly
consistent with the \citet{Rudie2012} measurement, shown by the red
dashed line, although note again that from Fig.~\ref{fig:Trho} this
should be expected.  The blue line displays the lower cut-off measured
from the mock data in the same way as \citet{Rudie2012}.  A column
density of $N_{\rm HI}\simeq 10^{12.95}\rm\,cm^{-2}$ corresponds to
the mean background density at this redshift \citep{Bolton2014},
suggesting that the value of $T_{0}=11735\rm\,K$ used in the 40-2048
simulation is in reasonable agreement with the observations. On the
other hand, the slightly steeper slope relative to the red dashed line
indicates that reducing the value of $\gamma-1=0.57$ used in the
simulation by $\Delta \gamma \sim 0.05$--$0.1$ may provide better
agreement with the observational data at $z=2.4$.  As discussed in
\citet{Bolton2014} and Rorai et al. (2016, submitted), however, this
agreement does not rule out the possibility that underdense gas in the
IGM is hotter than expected from a simple extrapolation of this
temperature-density relation to $\Delta<1$.  The \Lya forest
absorption lines at $z<3$ typically probe mildly overdense gas
\citep{Becker2011,Lukic2015}.  Hot gas can still persist at
$\Delta<1$, allowing for consistency with the lower envelope of the
$b_{\rm HI}$--$N_{\rm HI}$ plane for \Lya absorption lines at $z=2.4$.
We stress the PDF (e.g. Rorai et al. 2016, submitted) and power
spectrum of the transmitted flux do not yet rule out this possibility.


\section{Conclusions} \label{sec:conclude}

In this work we introduce a new set of large scale, high resolution
hydrodynamical simulations of the IGM -- the Sherwood simulation
suite.  We perform a detailed comparison to high resolution ($R\simeq
40000$), high signal-to-noise ($\rm S/N\sim 50$) observations of the
\Lya forest over the redshift range $2 \leq z \leq 5$.  Our
conclusions are as follows:

\begin{itemize}

\item The observed effective optical depth at $2<z<2.5$ and $4<z<5$ is
  overpredicted by the recent \citet{HaardtMadau2012} ionising
  background model \citep[see also][]{Puchwein2015}.  The \HI
  photo-ionisation rate, $\Gamma_{\rm HI}$, in the
  \citet{HaardtMadau2012} model must be increased by
  $[87,\,3,\,47,\,65]$ per cent at $z=[2,\,3,\,4,\,5]$ in our
  reference 40-2048 simulation, corresponding to $\Gamma_{\rm
    HI}=[1.76,\,0.86,\,0.83,\,0.71]\times 10^{-12}\rm\,s^{-1}$, in
  order to match the effective optical depth evolution described by
  Eq.~(\ref{eq:taueff}).  An in depth analysis of the \HI
  photo-ionisation rate that includes systematic uncertainties is
  presented in \citet{BeckerBolton2013}.
  
\item The observed transmitted flux PDF from \citet{Kim2007} and
  \citet{Rollinde2013} at $F=0$ is not recovered correctly in the
  simulations at $z=2.5$.  This is most likely due to the uncertain
  effect of star formation and galactic outflows, which increase the
  incidence of saturated ($N_{\rm HI}>10^{14.5}\rm\,cm^{-2}$) \Lya
  absorption lines, as well as the detailed signal-to-noise properties
  of the spectra \citep{Kim2007}.  The observed PDF at $0.1<F<0.8$
  furthermore still lies systematically below the mock data, although
  we find it is generally within the $2\sigma$ uncertainty we estimate
  from the models \citep[see also][]{Rollinde2013}.  This agreement
  may be improved for models with hotter underdense gas
  \citep{Bolton2014}.  A recent analysis of an ultra-high resolution
  quasar spectrum by Rorai et al. (2016, submitted) is consistent with
  this possibility.

\item We find that star formation and galactic winds have the largest
  impact on the \Lya forest at redshifts $z<2.5$ \citep[see
    also][]{Theuns2002,Viel2013feedback}.  When compared to
  simulations optimised for \Lya forest modelling which simply convert
  all gas with $\Delta>1000$ and $T<10^{5}\rm \,K$ into stars, the
  \citet{PuchweinSpringel2013} star formation and winds sub-grid model
  produces a greater incidence of saturated \Lya absorption systems
  and increases the transmitted flux power spectrum on large scales.
  At $z\geq 4$, however, this has little impact on the \Lya forest.

\item The effect of later reionisation (and hence less pressure
  smoothing) and the suppression of small scale power by a warm dark
  matter thermal relic has a large impact on the \Lya forest at $z>4$.
  Decreasing the pressure smoothing scale acts on the PDF and power
  spectrum in the opposite direction to warm dark matter.  It produces
  more pixels with $F>0.6$ in the PDF, and increases the power
  spectrum at scales $\log(k/\rm s\,km^{-1})>-1.5$.  This indicates it
  should be possible to constrain the integrated thermal history
  during reionisation using the line of sight \Lya forest power
  spectrum at high redshift \citep{Nasir2016}.

\item Mock \Lya forest spectra extracted from hydrodynamical
  simulations are in better agreement with the observed PDF and power
  spectrum if a correction for the uncertain continuum normalisation
  is applied.  The continuum is typically placed too low on the
  observed spectra, and the correction varies from a few per cent at
  $z=2$ to as much as $20$ per cent at $z=5$ \citep[see
    also][]{FaucherGiguere2008}.  This changes the shape of the PDF at
  $F>0.8$ and increases the power spectrum on all scales.

\item The observed scatter in the mean transmitted flux
  \citep{FaucherGiguere2008} is in reasonable agreement with the
  simulations at $z<3$.  However, this quantity is sensitive to the
  continuum placement on the simulated spectra.  Increasing
  (decreasing) the continuum level decreases (increases) the scatter
  in the mean transmission.  Variations in this correction may explain
  the broader scatter in the data at $z>3$ relative to the
  simulations.

\item The simulations are in good agreement with the CDDF of \Lya
  forest absorbers presented by \citet{Kim2013} at $2<z<3$.  The only
  discrepancies are that the simulations underpredict the number of
  weak lines with $N_{\rm HI} = 10^{12.4}\rm\,cm^{-2}$, and (at
  $z=2.1$ only) underpredict the incidence of saturated absorption
  lines with $N_{\rm HI}>10^{14.5}\rm cm^{-2}$.  We suggest the former
  may be due to unidentified metals and the detailed signal-to-noise
  properties of the data, whereas the latter agreement is improved
  (but not resolved) by including a sub-grid model for star formation
  and galactic winds.

\item The observed distribution of \Lya absorption line velocity
  widths \citep{Kim2013} is in good overall agreement with the
  simulations at $z=2.1$ and $z=2.7$, although absorption lines with
  $b\leq 10 \rm km\,s^{-1}$ are underpredicted by the models.  This
  difference is likely due to the presence of unidentified narrow
  metal lines in the observational sample.  At $z=2.7$, the
  simulations slightly overpredict the number of lines with $b_{\rm
    HI}=17$--$20\rm\,km\,s^{-1}$.  This suggests the simulations may
  be slightly too cold at $z=2.7$, possibly due to additional
  non-equilibrium heating not included in the simulations
  \citep{Puchwein2015}.

\item The lower cut-off in the $b_{\rm HI}$--$N_{\rm HI}$ distribution
  for lines with $10^{12.5}\rm\,cm^{-2}\leq N_{\rm HI}\leq
  10^{14.5}\rm\,cm^{-2}$ measured by \citet{Rudie2012} and reanalysed
  by \citet{Bolton2014} at $z=2.4$ is in broad agreement with our
  reference simulation, which has a temperature at mean density of
  $T_{0}=11735\rm\,K$ at this redshift.  However, a slightly lower
  value (by $\Delta \gamma \simeq 0.05$--$0.1$) for the slope of
  temperature-density relation assumed in the models at this redshift,
  $\gamma-1=0.57$, may provide better agreement with the observational
  data at $z=2.4$.  We stress, however, that \Lya forest absorption
  lines at $z<3$ typically probe mildly overdense gas.  Hot gas may
  still persist at $\Delta<1$ while still allowing consistency with
  the lower envelope of the $b_{\rm HI}$--$N_{\rm HI}$ plane for \Lya
  absorption lines at $z=2.4$ (see Rorai et al. 2016, submitted).

\end{itemize}

\noindent
Overall, we conclude the Sherwood simulations are in very good
agreement with a wide range of \Lya forest data at $2<z<5$. These
results lend further support to the now well established paradigm that
the \Lya forest is a natural consequence of the web-like distribution
of matter arising in \LCDM cosmological models.  However, a number of
small discrepancies still remain with respect to the observational
data, motivating further observational and theoretical investigation.
We suggest that in the short-term, improved measurements of the power
spectrum and PDF at $z>3$ using larger, high resolution data sets, and
tighter constraints on the slope of the temperature-density relation
at $z \simeq 3$ will be particularly beneficial.

\section*{Acknowledgments}

The hydrodynamical simulations used in this work were performed with
supercomputer time awarded by the Partnership for Advanced Computing
in Europe (PRACE) 8th Call. We acknowledge PRACE for awarding us
access to the Curie supercomputer, based in France at the Tr{\'e}
Grand Centre de Calcul (TGCC).  This work also made use of the DiRAC
High Performance Computing System (HPCS) and the COSMOS shared memory
service at the University of Cambridge. These are operated on behalf
of the STFC DiRAC HPC facility.  This equipment is funded by BIS
National E-infrastructure capital grant ST/J005673/1 and STFC grants
ST/H008586/1, ST/K00333X/1.  We thank Volker Springel for making
\textsc{P-GADGET-3} available. JSB acknowledges the support of a Royal
Society University Research Fellowship.  MGH and EP acknowledge
support from the FP7 ERC Grant Emergence-320596, and EP gratefully
acknowledges support by the Kavli Foundation. DS acknowledges support
by the STFC and the ERC starting grant 638707 ``Black holes and their
host galaxies: co-evolution across cosmic time''.  JAR is supported by
grant numbers ST/L00075X/1 and RF040365.  MV and TSK are supported by
the FP7 ERC grant ``cosmoIGM'' and the INFN/PD51 grant.


\appendix
\section{Numerical convergence}

\begin{figure}
\begin{center}
  \includegraphics[width=0.47\textwidth]{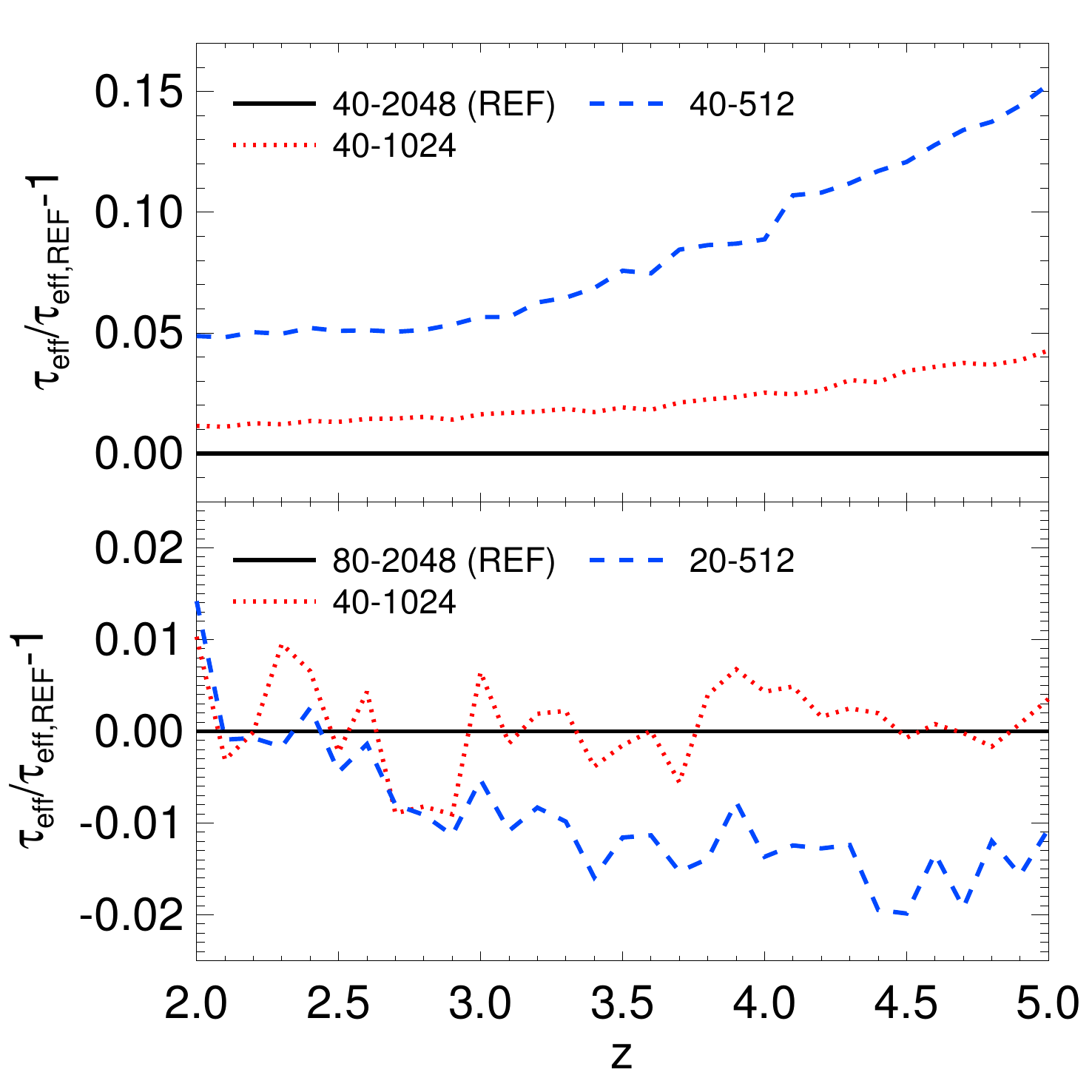}
  \vspace{-0.4cm}
  \caption{The redshift evolution of the effective optical depth
    relative to the reference model indicated in each panel.  {\it
      Top:} Convergence with mass resolution for a fixed box size of
    $40h^{-1}\,\rm cMpc$. {\it Bottom:} Convergence with box size for
    a fixed mass resolution of $M_{\rm gas}=7.97\times 10^{5}
    h^{-1}\,M_{\odot}$. Note that $\tau_{\rm eff}$ has not been
    rescaled in this comparison.}
  \label{fig:taueff_conv}
\end{center}
\end{figure}

\begin{figure}
\begin{center}
  \includegraphics[width=0.47\textwidth]{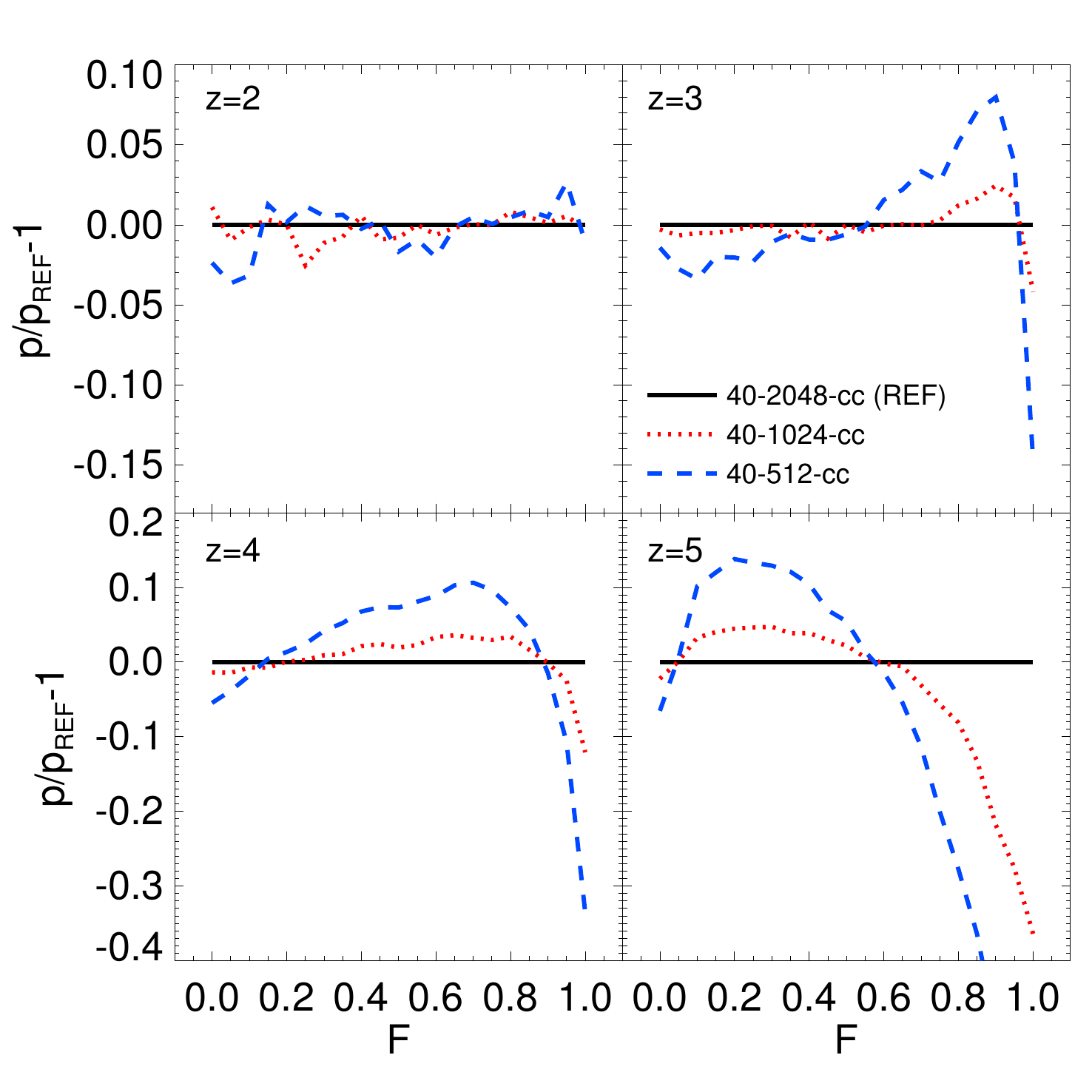}
  \vspace{-0.4cm}
  \caption{The probability distribution of the transmitted flux at
    $z=2,\,3,\,4$ and $5$ relative to the reference model indicated in
    the upper right panel.  Each panel displays the convergence with
    mass resolution for a fixed box size of $40h^{-1}\,\rm cMpc$ at
    the four different redshifts. Note that $\tau_{\rm eff}$ has been
    rescaled to match Eq.~(\ref{eq:taueff}) and the mock spectra are
    convolved with a $7\rm\,km\,s^{-1}$ Gaussian instrument
    profile.}
  \label{fig:pdf_res}
\end{center}
\end{figure}

\begin{figure}
\begin{center}
  \includegraphics[width=0.47\textwidth]{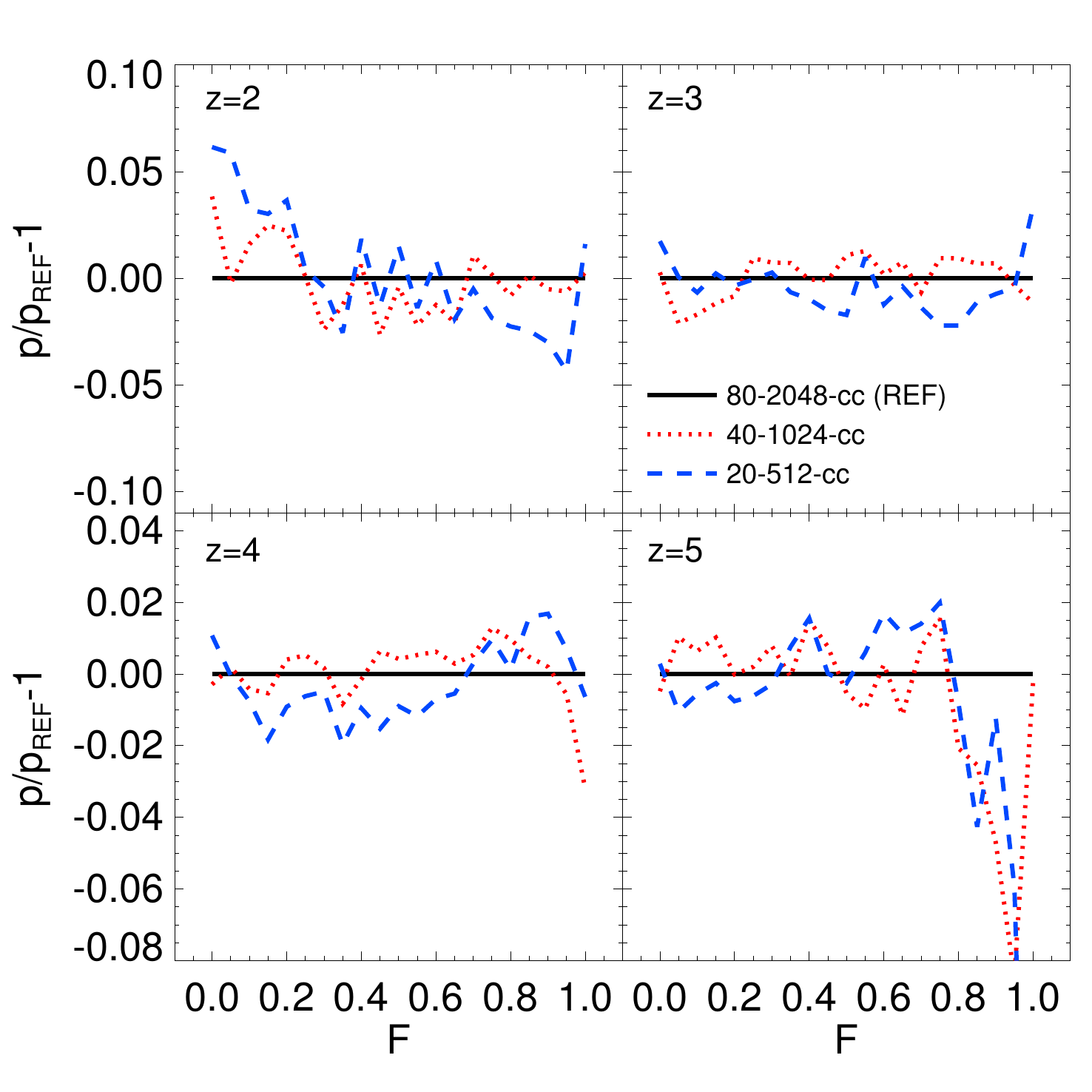}
  \vspace{-0.4cm}
  \caption{The probability distribution of the transmitted flux at
    $z=2,\,3,\,4$ and $5$ relative to the reference model indicated in
    the upper right panel.  Each panel displays the convergence with
    box size for a fixed mass resolution of $M_{\rm gas}=7.97\times
    10^{5} h^{-1}\,M_{\odot}$ at the four different redshifts.  Note
    that $\tau_{\rm eff}$ has been rescaled to match
    Eq.~(\ref{eq:taueff}) and the mock spectra are convolved with a
    $7\rm\,km\,s^{-1}$ Gaussian instrument profile.}
  \label{fig:pdf_box}
\end{center}
\end{figure}

\begin{figure}
\begin{center}
  \includegraphics[width=0.47\textwidth]{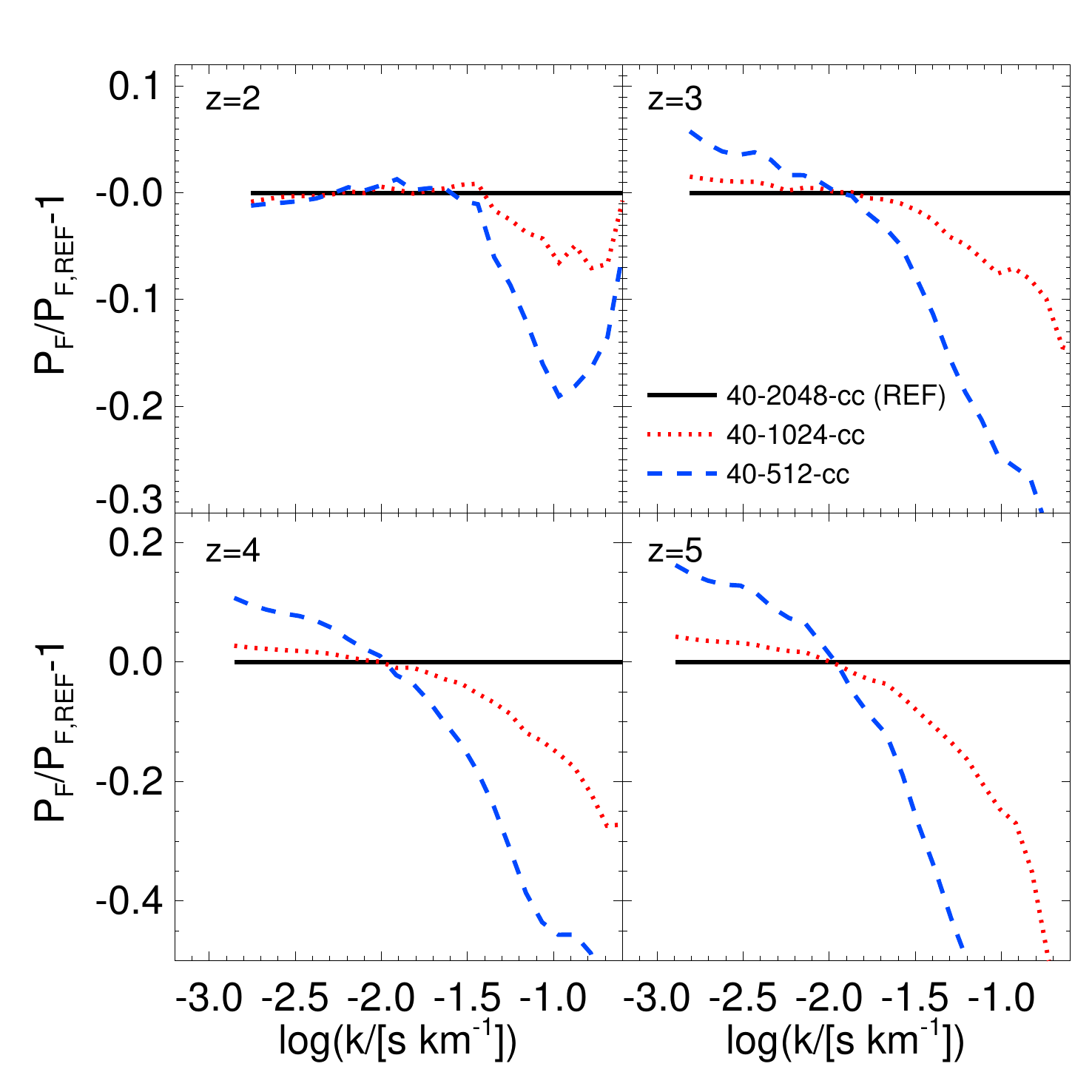}
  \vspace{-0.4cm}
  \caption{The power spectrum of the transmitted flux at $z=2,\,3,\,4$
    and $5$ relative to the reference model indicated in the upper
    right panel.  Each panel displays the convergence with mass
    resolution for a fixed box size of $40h^{-1}\,\rm cMpc$ at the
    four different redshifts.  Note that $\tau_{\rm eff}$ has been
    rescaled to match Eq.~(\ref{eq:taueff}) and the mock spectra are
    convolved with a $7\rm\,km\,s^{-1}$ Gaussian instrument
    profile for all models.}
  \label{fig:pk_res}
\end{center}
\end{figure}

\begin{figure}
\begin{center}
  \includegraphics[width=0.47\textwidth]{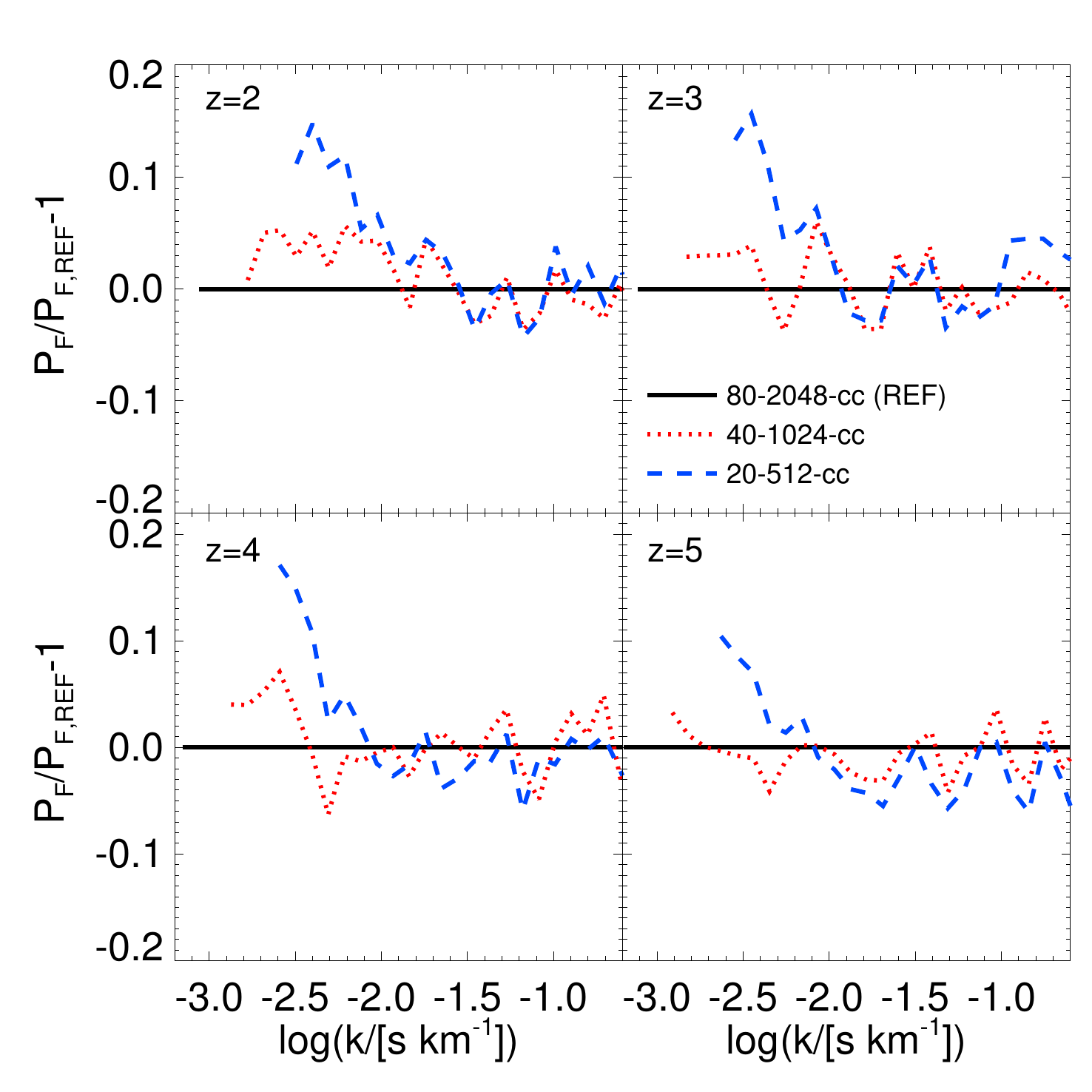}
  \vspace{-0.4cm}
  \caption{The power spectrum of the transmitted flux at $z=2,\,3,\,4$
    and $5$ relative to the reference model indicated in the upper
    right panel.  Each panel displays the convergence with box size
    for a fixed mass resolution of $M_{\rm gas}=7.97\times 10^{5}
    h^{-1}\,M_{\odot}$ at the four different redshifts.  Note that
    $\tau_{\rm eff}$ has been rescaled to match Eq.~(\ref{eq:taueff})
    and the mock spectra are convolved with a $7\rm\,km\,s^{-1}$
    Gaussian instrument profile.}
  \label{fig:pk_box}
\end{center}
\end{figure}

\begin{figure}
\begin{center}
  \includegraphics[width=0.47\textwidth]{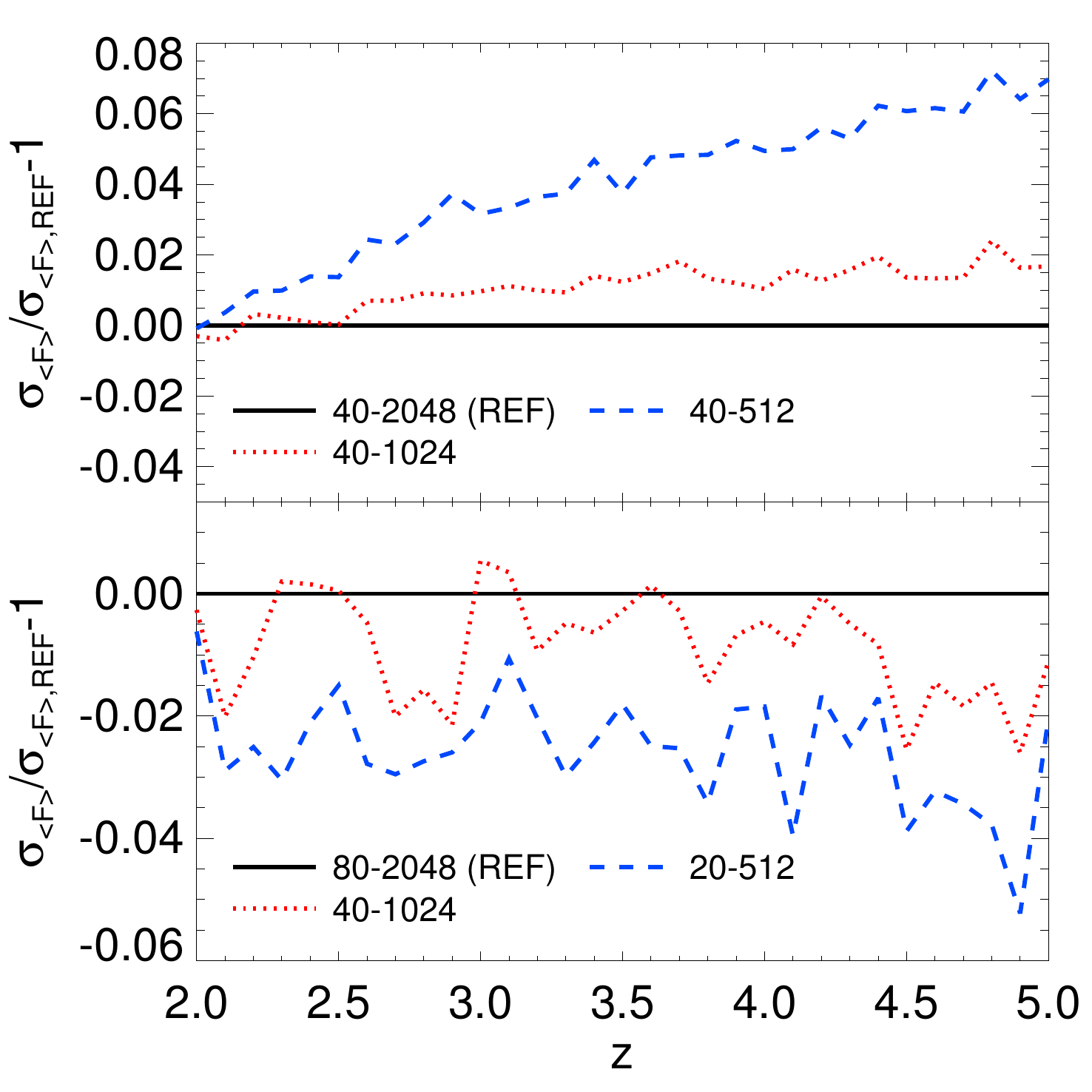}
  \vspace{-0.4cm}
  \caption{The redshift evolution of the standard deviation of the
    mean transmission in $3\rm\,pMpc$ segments relative to the
    reference model indicated in each panel.  {\it Top:} Convergence
    with mass resolution for a fixed box size of $40h^{-1}\,\rm
    cMpc$. {\it Bottom:} Convergence with box size for a fixed mass
    resolution of $M_{\rm gas}=7.97\times 10^{5} h^{-1}\,M_{\odot}$.
    Note that $\tau_{\rm eff}$ has been rescaled to match
    Eq.~(\ref{eq:taueff}).}
  \label{fig:sigma_conv}
\end{center}
\end{figure}

\begin{figure}
\begin{center}
  \includegraphics[width=0.47\textwidth]{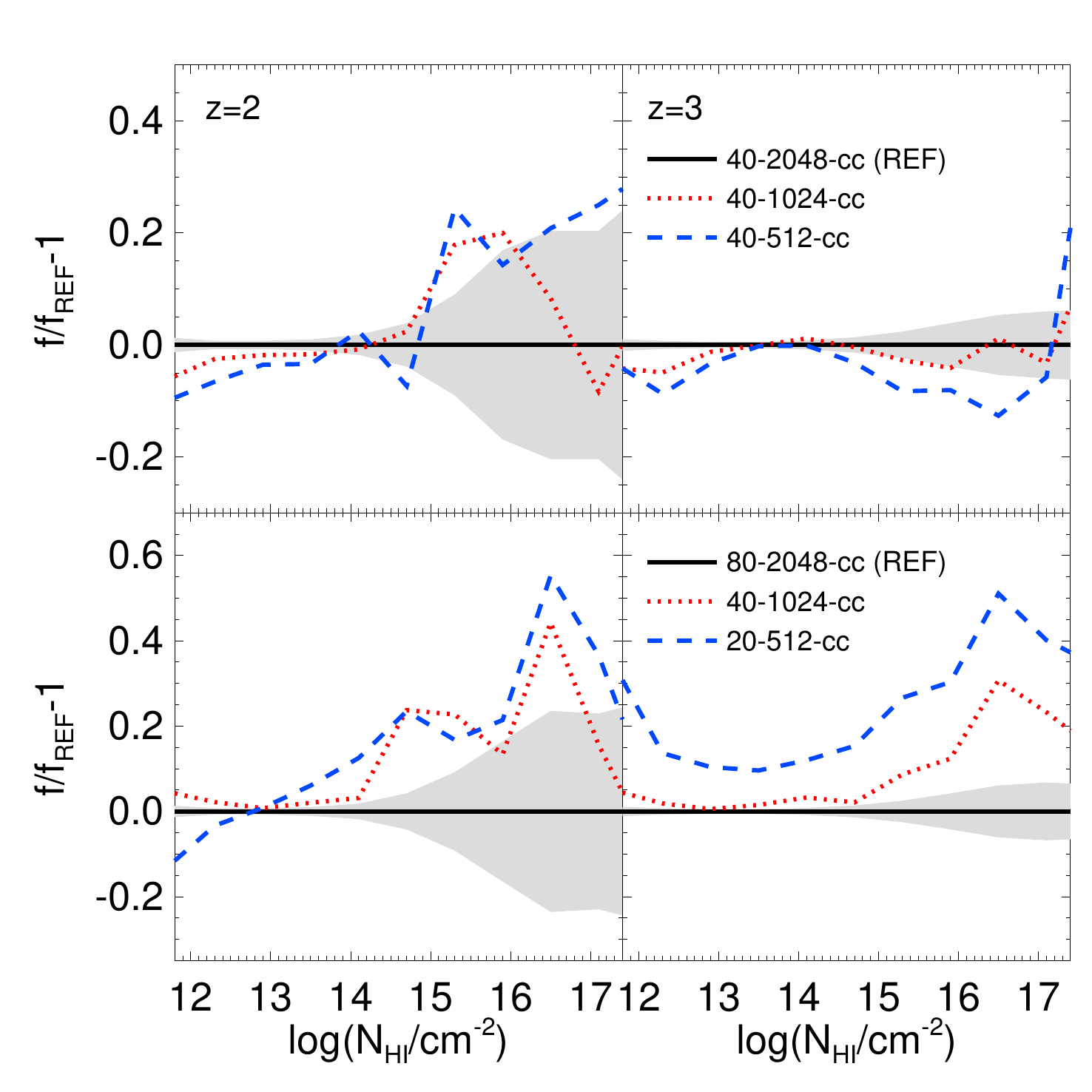}
  \vspace{-0.4cm}
  \caption{The column density distribution function at $z=2$ (left
    column) and $z=3$ (right column) relative to the reference models
    indicated in each row.  {\it Top panels:} Convergence with mass
    resolution for a fixed box size of $40h^{-1}\,\rm cMpc$. {\it
      Bottom panels:} Convergence with box size for a fixed mass
    resolution of $M_{\rm gas}=7.97\times 10^{5}
    h^{-1}\,M_{\odot}$. Note that $\tau_{\rm eff}$ has been rescaled
    to match Eq.~(\ref{eq:taueff}) for all models and the mock spectra
    have been processed as described in Section~\ref{sec:spectra}. The
    grey shading corresponds to the Poisson error in each bin.}
  \label{fig:cddf_conv}
\end{center}
\end{figure}

\begin{figure}
\begin{center}
  \includegraphics[width=0.47\textwidth]{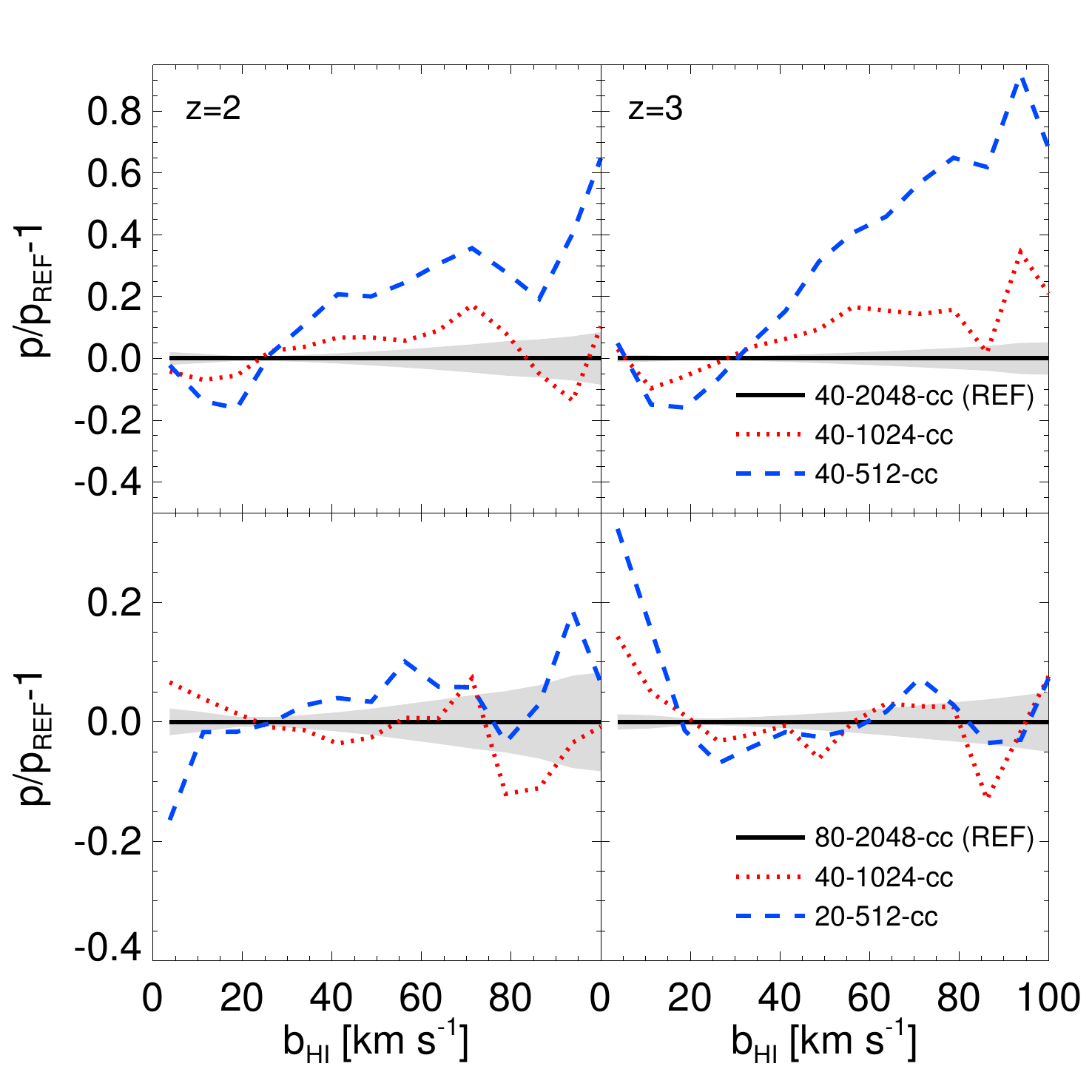}
  \vspace{-0.4cm}
  \caption{The velocity width distribution of \Lya absorption lines at
    $z=2$ (left column) and $z=3$ (right column) relative to the
    reference models indicated in each row.  {\it Top panels:}
    Convergence with mass resolution for a fixed box size of
    $40h^{-1}\,\rm cMpc$. {\it Bottom panels:} Convergence with box
    size for a fixed mass resolution of $M_{\rm gas}=7.97\times 10^{5}
    h^{-1}\,M_{\odot}$. Note that $\tau_{\rm eff}$ has been rescaled
    to match Eq.~(\ref{eq:taueff}) for all models and the mock spectra
    have been processed as described in Section~\ref{sec:spectra}. The
    grey shading corresponds to the Poisson error in each bin.}
  \label{fig:bdist_conv}
\end{center}
\end{figure}

Fig.~\ref{fig:taueff_conv}--\ref{fig:bdist_conv} display convergence
tests with mass resolution and box size for all the quantities studied
in this paper.  All results are computed from mock spectra with a
total path length of $10^{5}h^{-1}\rm\,cMpc$.  In general, we find
that the convergence properties of the simulations are excellent;
simulation volumes that are $40h^{-1}\rm\,cMpc$ on a side with a gas
particle mass of $7.97\times 10^{5}h^{-1}M_{\odot}$ are sufficient for
resolving many of the statistics examined in this work.  Similar
results along with a more detailed discussion of these issues may be
found in \citet{BoltonBecker2009} and more recently \citet{Lukic2015}.

All the transmitted flux statistics are converged to within $\sim 5$
per cent with respect to mass resolution and box size except at $z\geq
4$, where the PDF at $F>0.8$ and the power spectrum on scales
$\log(k/\rm s\,km^{-1})>-1$ are converged at $10$--$20$ per cent
level.  With regard to the Voigt profile fits, we find the CDDF at
$N_{\rm HI}<10^{14.5}\rm\,cm^{-2}$ and the velocity width distribution
are also converged to within $\sim 5$--$10$ per cent with box size and
mass resolution at $2 \leq z \leq 3$.  At $N_{\rm
  HI}>10^{14.5}\rm\,cm^{-2}$, the CDDF is converged at the $10$--$20$
per cent level, although note the variance in the CDDF can become
comparable to this toward high column densities as the number of lines
in each bin becomes very small.  This is illustrated by the grey
shading in Fig.~\ref{fig:cddf_conv} and Fig.~\ref{fig:bdist_conv},
corresponding to the Poisson error in each bin.

\end{document}